\PassOptionsToPackage{unicode}{hyperref}
\PassOptionsToPackage{hyphens}{url}

\documentclass[11pt,a4paper]{article}

\usepackage[T1]{fontenc}
\usepackage[utf8]{inputenc}
\usepackage{lmodern}

\usepackage{amsmath,amssymb}
\usepackage{bm}
\usepackage{tensor}
\usepackage{xcolor}
\usepackage{graphicx}

\usepackage[margin=1in]{geometry}

\usepackage[numbers,sort&compress]{natbib}
\usepackage{tabularx}
\usepackage{longtable,booktabs}
\usepackage{calc}

\newcommand{\pandocbounded}[1]{#1}

\AtBeginEnvironment{quote}{\color{red}}

\usepackage{hyperref}
\usepackage{bookmark}
\IfFileExists{xurl.sty}{\usepackage{xurl}}{}
\usepackage{orcidlink}

\hypersetup{
pdftitle={Probing Rotating Einstein–Power-Yang–Mills Black Holes through Shadows and Quasinormal Modes: Prospects for Event Horizon Telescope Constraints},
pdfkeywords={Black hole shadow \and Quasinormal modes \and Power-Yang-Mills field \and Event Horizon Telescope \and Newman-Janis algorithm},
pdfauthor={I. Ezzaki; A. El Boukili;  H. Lekbich; A. Benami},
pdfsubject={Probing Rotating Einstein–Power-Yang–Mills Black Holes through Shadows and Quasinormal Modes: Prospects for Event Horizon Telescope Constraints},
pdflang={en},
pdfdisplaydoctitle=true,
colorlinks=true,
linkcolor=red,
citecolor=blue,
urlcolor=blue,
breaklinks=true
}

\title{\textbf{Probing Rotating Einstein–Power-Yang–Mills Black Holes through Shadows and Quasinormal Modes: Prospects for Event Horizon Telescope Constraints}}

\author{
I. Ezzaki$^{1,2}$\orcidlink{0009-0005-3476-1873}\thanks{is.ezzaki@edu.umi.ac.ma}
\and
A. El Boukili$^{1}$\orcidlink{0000-0002-3277-9640}\thanks{a.elboukili@umi.ac.ma}
\and 
H. Lekbich$^{1}$\orcidlink{0009-0009-6079-0394}\thanks{h.lekbich@edu.umi.ac.ma}
\and
A. Benami$^{2}$\orcidlink{0000-0001-5516-5660}
}

\date{}

\begin{document}

\maketitle

\begin{center}
\small
$^{1}$ PMRA team, SIMED Laboratory, Faculty of Sciences and Techniques,
Errachidia, Moulay Ismail University of Meknes, Morocco

\vspace{0.3cm}

$^{2}$ OTEA team, SIMED Laboratory, Faculty of Sciences and Techniques,
Errachidia, Moulay Ismail University of Meknes, Morocco
\end{center}

\vspace{0.5cm}

\maketitle

\begin{abstract}
Within Einstein power-Yang-Mills gravity (EPYM), we build a rotating black hole via the Newman-Janis procedure applied to the spherically symmetric static seed, obtaining an EPYM-inspired Kerr-like metric controlled by the spin \(a\), the magnetic Yang-Mills charge \(Q\), and the power parameter \(q\).
We analyse the horizon structure and the photon region, and we compute the shadow as it appears to a remote observer for different values of these parameters.
Using the published Event Horizon Telescope measurements of M87\(^*\) and Sgr A\(^*\), we constrain the Yang-Mills charge and the power parameter from the angular size and the Schwarzschild deviation \(\delta\) of the observed images.
At the Maxwell point \(q = 1\) the spin marginalized likelihood of M87\(^*\) bounds the charge to \(Q \lesssim 0.29\,M\) at \(1\sigma\), while the fixed-spin band intersection at \(a = 0.7\,M\) gives the weaker \(Q \lesssim 0.52\,M\) from the M87\(^*\) angular size alone. The bound weakens as \(q \to 3/2\), where the shadow diameter becomes nearly insensitive to the charge.
The shadow radius also fixes the limiting absorption cross section at high frequencies and hence the energy emission rate, which is suppressed by the charge and enhanced by the power parameter.
We then compute the quasinormal modes of a massless scalar on the rotating background from the leading-order Wentzel-Kramers-Brillouin (WKB) conditions on the Teukolsky type radial potential, finding that the oscillation frequency rises and the damping rate falls as the charge or spin increases.
The power parameter leaves an imprint on the shadow size and its charge sensitivity and in the quasinormal spectrum the shift is in principle present but lies below the resolution of current detectors.
\end{abstract}
\subsection*{keywords:}Black hole shadow \and Quasinormal modes \and Power-Yang-Mills field \and Event Horizon Telescope \and Newman-Janis algorithm
\newpage

\section{Introduction}\label{sec:intro}

The Event Horizon Telescope (EHT) images of the supermassive objects at the centres of M87\(^*\) and the Milky Way moved strong-field gravity from a theoretical arena into a measurement one \citep{EventHorizonTelescope:2019dse, EventHorizonTelescope:2019ggy, EventHorizonTelescope:2022wkp, EventHorizonTelescope:2022xqj}.
The bright ring and the central brightness depression seen in those images are set by the photon region of the underlying spacetime, so the angular size and the form of the shadow now act as observational probes of the metric.
A black hole is also an oscillator.
When it is perturbed it rings down through a discrete set of damped quasinormal modes (QNMs), and in the eikonal limit these modes are tied to the same unstable photon orbits that bound the shadow \citep{Cardoso:2008bp, Jusufi:2019ltj}.
Shadow and ringdown probe the geometry at overlapping radii, which makes them a natural pair to compute for any candidate metric.

The idea that the black hole shadow constitutes a direct image of the photon capture sphere goes back to the early work of Synge \citep{Synge:1966okc} and Luminet \citep{Luminet:1979nyg} for static metrics
and Bardeen for the rotating case \citep{Bardeen:1973tla}.
Falcke et al. were the first to propose observing this feature for Sgr A\(^*\) \citep{Falcke:1999pj}, and the EHT collaboration confirmed the concept in 2019 through a dedicated global very long baseline interferometry (VLBI) array and a reconstructed image of the M87\(^*\) ring \citep{EventHorizonTelescope:2019uob, EventHorizonTelescope:2019ths}.
The photon sphere radius is linked directly to the apparent shadow size seen at infinity, so comparing the predicted shadow angular diameter against the EHT measurements bounds the parameters of any candidate metric family \citep{Vagnozzi:2022moj, Kumar:2020owy, Perlick:2021aok, Afrin:2021imp, Afrin:2021wlj, KumarWalia:2022aop, Pantig:2022qak, Ghorani:2023hkm}.

The connection between the shadow and quasinormal modes follows from the same photon orbit.
In the eikonal regime the oscillatory part of the frequency tracks the angular velocity \(\Omega_c\) of that unstable orbit, while the decaying part is governed by the Lyapunov exponent \(\lambda_L\) that characterises how quickly perturbations escape the vicinity of the orbit \citep{Cardoso:2008bp, Jusufi:2019ltj, Konoplya:2017wot}.
For rotating spacetimes the shadow becomes asymmetric, and the generalisation of the correspondence to the axisymmetric case, explored in detail by Yang \citep{Yang:2021zqy} and applied to rotating regular black holes by Pedrotti and Vagnozzi \citep{Pedrotti:2024znu}, requires that both the Hamilton-Jacobi and the Klein-Gordon equations separate.
Shadow size and ringdown frequencies are set by the same photon orbits and together form a self-consistent diagnostic of the near-horizon spacetime that motivates the program of this paper.

Nonlinear gauge fields offer a natural way to deform the Schwarzschild and Kerr geometries while retaining a self-contained Lagrangian description.
The power-Yang-Mills field, in which the standard Yang-Mills invariant \(\mathcal{F}\) is raised to a power \(q\), generalises the Maxwell and Yang-Mills cases through a single dimensionless exponent \citep{Mazharimousavi:2009mb}.
In the resulting Einstein-power-Yang-Mills, or EPYM, black holes carry a magnetic charge \(Q\) together with the power \(q\), and they reduce to the Reissner-Nordström solution when \(q=1\).
Their thermodynamics, shadows, and quasinormal spectra have been examined in the non-rotating, spherically symmetric case \citep{Biswas:2022qyl, Zubair:2023cep, Gogoi:2023ffh, Rincon:2023hvd}, where the power parameter is found to shift the horizon and the photon sphere in a way that the charge alone does not capture.

Nonlinear electrodynamics (NED) coupled to general relativity has attracted sustained attention since Born and Infeld introduced it to regularize the Coulomb singularity in 1934 \citep{Born:1934gh}.
Extensions motivated by string theory and quantum field theory, among them the Euler-Heisenberg model arising from one-loop vacuum polarisation \citep{Heisenberg:1936nmg}, produce black hole geometries that differ appreciably from the Reissner-Nordström solution in the strong-field region, among them the regular NED black holes of Ayón-Beato and García \citep{Ayon-Beato:1998hmi, Ayon-Beato:1999kuh, Ayon-Beato:1999qin, Ezzaki:2025abg}.
The Yang-Mills field occupies a special place among non-Abelian gauge sources because its coupling to Einstein gravity admits soliton and hairy black hole solutions that are impossible when the two sectors are treated separately.
The initial motivation for coupling power-Yang-Mills fields to gravity, studied in the context of Lovelock black holes by Mazharimousavi and Halilsoy \citep{Mazharimousavi:2009mb, HabibMazharimousavi:2008ib}, was the construction of exact solutions whose thermodynamics exhibit van der Waals-like transitions, Joule-Thomson expansion, and other features that the linear Yang-Mills sector lacks \citep{Biswas:2022qyl}.

Subsequent work on Einstein gravity with both Maxwell and power-Yang-Mills sources showed that combining Abelian and non-Abelian charges generates a richer parameter space, allowing the shadow radius and quasinormal spectrum to vary with the Yang-Mills charge in a way that can in principle be disentangled from purely electromagnetic effects \citep{Rincon:2023hvd, Gogoi:2023ffh}.
The power exponent \(q\) and the Yang-Mills charge \(Q\) affect observables in opposite directions, as demonstrated in the static four-dimensional case \citep{Gogoi:2023ffh, Chakhchi:2022fls}, when we increase the charge \(Q\) the shadow shrinks and raises the real quasinormal frequency, while increasing \(q\) enlarges the shadow and reduces the oscillation frequency.
This opposite sensitivity makes EPYM black holes particularly attractive as test beds, since the two parameters can in principle be constrained independently by combining shadow size, ringdown frequency, and emission rate.

The static, spherically symmetric EPYM black hole in four dimensions has been studied from several complementary angles.
Its thermodynamic phase structure in anti-de Sitter space, including van der Waals-like transitions and Joule-Thomson cooling, was investigated by Biswas \citep{Biswas:2022qyl} and extended to higher dimensions by Zubair et al. \citep{Zubair:2023cep}.
The quasinormal modes of the four-dimensional solution for the scalar perturbation were computed using the Wentzel-Kramers-Brillouin (WKB) method by Gogoi et al. \citep{Gogoi:2023ffh}, who established the QNM-shadow correspondence in the eikonal limit and found that the Yang-Mills charge and the power exponent produce opposite shifts in the oscillation frequency and in the damping rate.
Rincon and collaborators studied the Einstein-Maxwell-power-Yang-Mills black hole at the fractional power \(p=1/2\), which carries both Abelian and non-Abelian charges, computed its scalar quasinormal spectrum with the WKB method and verified its stability, and bounded the two charges with the Sgr A\(^*\) shadow \citep{Rincon:2023hvd}.
The optical appearance of a power-Yang-Mills black hole surrounded by thin accretion disk models was studied by Chakhchi et al. \citep{Chakhchi:2022fls}, who used backward ray-tracing to separate the direct emission from the lensing and photon ring contributions and showed that the shadow area decreases with increasing Yang-Mills charge.
Luo et al. \citep{Luo:2025ocw} traced the photon rings of the same Einstein-Maxwell-power-Yang-Mills black hole at \(p=1/2\) by backward ray-tracing and found that the horizon, the photon sphere, and the shadow area all shrink as the Yang-Mills hair grows. In a related higher-dimensional direction, Yan \citep{Yan:2025gks} investigated EPYM black holes within the Gauss-Bonnet framework, finding that the Gauss-Bonnet coupling dominates over both the Yang-Mills charge and the power exponent in modifying the shadow and perturbation spectra.
Among these works only Zubair et al. \citep{Zubair:2023cep} consider a rotating metric, and there the power-Yang-Mills charge acts alongside a Gauss-Bonnet coupling. The combined effect of spin, charge, and power-law exponent on the shadow observables and quasinormal spectrum of the pure EPYM theory remains unexplored.

What is missing is the rotating case confronted with data.
Astrophysical black holes spin, and the EHT targets are no exception, so a static EPYM model cannot be matched to the observed images in a controlled way.
The closest antecedent is the rotating power-Yang-Mills black hole with a Gauss-Bonnet coupling, whose shadow was confronted with both EHT targets in \citep{Zubair:2023cep}. To our knowledge, no rotating extension of the pure EPYM black hole, free of a Gauss-Bonnet term, has yet been built and then constrained against the measured shadow sizes of M87\(^*\) and Sgr A\(^*\) \citep{Gogoi:2023ffh, Chakhchi:2022fls}, nor has the ringdown of any rotating power-Yang-Mills geometry been computed alongside the shadow within one framework.

Generating an exact rotating black hole solution from a non-rotating seed is technically demanding, and most of the rotating metrics studied in the literature were not derived by directly solving the full set of Einstein equations with a spinning ansatz.
The Newman-Janis algorithm offers a systematic route.
In its original form \citep{Newman:1965tw, Newman:1965my} the algorithm involves a complex coordinate transformation applied to the static metric in null coordinates, followed by a complexification procedure whose ambiguities can lead to inconsistent rotating metrics when the source is not vacuum, although the algorithm nonetheless reproduces the Kerr-Newman metric uniquely from its static seed \citep{Drake:1998gf}.
Azreg-Aïnou removed this ambiguity by formulating a complexification-free version of the algorithm, which generates a rotating metric without introducing unphysical cross-terms and reproduces the Kerr-Newman solution exactly in the appropriate limit \citep{Azreg-Ainou:2014aqa, Azreg-Ainou:2014pra}.
This is the version we apply here.

Shaikh showed that for the broad class of rotating metrics that the Newman-Janis algorithm generates, the null geodesic equations are always separable and a general analytic formula for the shadow contour can be derived \citep{Shaikh:2019fpu}.
That result means the shadow computation is valid for our rotating EPYM metric without additional assumptions.
The rotating metrics obtained from this complexification-free route also satisfy the separability conditions for the Klein-Gordon equation identified by Chen and Chen \citep{Chen:2019jbs}, which are the conditions required for the eikonal QNM-shadow correspondence to carry over from the Kerr case to the metric under study \citep{Vagnozzi:2022moj, Yang:2021zqy}.
The algorithm is most reliable for vacuum and near vacuum solutions. When an external non-Abelian field sources the geometry, the resulting metric need not solve the original field equations exactly, and we adopt it here in the phenomenological spirit made explicit inSec.~\ref{sec:solution}.

We proceed in three steps.
First, we generate the rotating EPYM metric by the complexification-free Newman-Janis algorithm applied to the static seed \citep{Shaikh:2019fpu, Azreg-Ainou:2014aqa, Azreg-Ainou:2014pra} and examine its horizons and photon region (Sec.~\ref{sec:solution}).
Second, we compute the shadow as it appears to a remote observer and extract the coordinate-independent observables, confront the angular size and the Schwarzschild deviation \(\delta\) with the M87\(^*\) and Sgr A\(^*\) EHT measurements to bound \(Q\) and \(q\), sharpening the result with a spin marginalized likelihood \citep{Kumar:2020owy, Vagnozzi:2022moj} (Sec.~\ref{sec:shadow},Sec.~\ref{sec:eht}), and derive the energy emission rate from the shadow radius together with the Hawking temperature (Sec.~\ref{sec:shadow}).
Finally, we obtain the quasinormal modes of a massless scalar field on the rotating background from the leading-order WKB treatment of the Teukolsky type radial equation \citep{Yang:2012he, Lambiase:2024lvo} (Sec.~\ref{sec:qnm}).

We work in geometrized units with \(G=c=\hbar=k_B=1\) and adopt the signature \((-,+,+,+)\).
The mass sets the scale, so the spin \(a\) and all radii are quoted in units of \(M\). Since the action Eq.~\ref{eq:action} carries no coupling in front of \(\mathcal{F}^q\), the charge has dimension \([Q]=L^{2-1/q}\). The metric is invariant, up to an overall factor \(\lambda^2\), under \((r,M,a,Q)\to(\lambda r,\lambda M,\lambda a,\lambda^{2-1/q}Q)\), so every dimensionless observable depends on the charge only through \(Q/M^{2-1/q}\), and we quote the charge in units of \(M^{2-1/q}\); at \(q=1\) this is the familiar \(Q/M\). Curves of different \(q\) at the same \(Q/M^{2-1/q}\) therefore compare equal dimensionless strength rather than equal physical charge, and where bounds are compared across \(q\) we also express them as fractions of the extremal charge \(Q_{\rm ext}\) at the same spin.
The exponent \(q\) is dimensionless and should not be read as an electric charge.

\section{Rotating EPYM black hole}\label{sec:solution}

We begin with the four-dimensional action of Einstein gravity coupled to a power-Yang-Mills source \citep{Mazharimousavi:2009mb, Biswas:2022qyl}

\begin{equation}\protect\phantomsection\label{eq:action}{I = \frac{1}{16\pi}\int d^4x \sqrt{-g}\,\big(\mathcal{R} - \mathcal{F}^{q}\big),}\end{equation}

where the exponent \(q\) is real and positive, \(\mathcal{R}\) denotes the Ricci scalar, and the Yang-Mills invariant \(\mathcal{F}=\mathrm{Tr}\big(\mathcal{F}^{(a)}_{\mu\nu}\mathcal{F}^{(a)\mu\nu}\big)\) is built from the gauge field strength \(\mathcal{F}^{(a)}_{\mu\nu}\).
The trace runs over the adjoint index \(a\) of the \(SO(3)\) gauge group appropriate to four dimensions, with three generators.
When \(q=1\) the source reduces to the ordinary Einstein-Yang-Mills theory \citep{HabibMazharimousavi:2007fst, Mazharimousavi:2008ap}, and the action Eq.~\ref{eq:action} is then the direct four-dimensional counterpart of the Lovelock-EPYM family studied in \citep{Mazharimousavi:2009mb}.

The Yang-Mills field strength \(\mathcal{F}^{(a)}_{\mu\nu}\) is built from the \(SO(3)\) connection \(A^{(a)}_\mu\),

\begin{equation}\protect\phantomsection\label{eq:sol-ym-field}{\mathcal{F}^{(a)}_{\mu\nu} = \partial_\mu A^{(a)}_\nu - \partial_\nu A^{(a)}_\mu + \frac{1}{2\varsigma}\,C^{(a)}_{(b)(c)}\,A^{(b)}_\mu A^{(c)}_\nu,}\end{equation}

where \(C^{(a)}_{(b)(c)}\) are the gauge Lie group structure constants and \(\varsigma\) is an arbitrary coupling constant.
To obtain a static, spherically symmetric solution one adopts the magnetic Wu-Yang ansatz \citep{Mazharimousavi:2008ap, Mazharimousavi:2009mb}

\begin{equation}\protect\phantomsection\label{eq:sol-wu-yang}{\mathbf{A}^{(a)} = \frac{Q}{r^2}\,\bigl(x_i\,dx_j - x_j\,dx_i\bigr),}\end{equation}

where the pair of Cartesian indices \((i,j)\) satisfies \(1\le j<i\le 3\), the gauge index \(a\) runs accordingly, and \(\mathbf{A}^{(a)} = A^{(a)}_\mu\,dx^\mu\) is the coordinate 1-form corresponding to the component potential \(A^{(a)}_\mu\) introduced in Eq.~\ref{eq:sol-ym-field}.
This purely magnetic configuration carries the Yang-Mills charge \(Q\), and the corresponding invariant evaluates to

\begin{equation}\protect\phantomsection\label{eq:sol-ym-inv}{\mathcal{F} = \mathrm{Tr}\bigl(\mathcal{F}^{(a)}_{\mu\nu}\mathcal{F}^{(a)\mu\nu}\bigr) = \frac{2Q^2}{r^4}.}\end{equation}

Note that \(\mathcal{F}\) is positive for all \(r>0\), so the power \(\mathcal{F}^q\) is real for any real \(q>0\).
The non-Abelian character of the gauge field is essential.
Under the Wu-Yang ansatz the \(SO(3)\) structure constants drop out of the field equation for \(f(r)\), and the invariant \(\mathcal{F}=2Q^2/r^4\) takes the same functional form as a power-Maxwell invariant.
Consequently the static lapse coincides with the power-Maxwell result even though the underlying gauge field remains genuinely non-Abelian \citep{Mazharimousavi:2009mb}.

The variation of Eq.~\ref{eq:action} with respect to \(g_{\mu\nu}\) yields the Einstein equations \(G^\mu{}_\nu = 8\pi T^\mu{}_\nu\), in which the stress tensor of the power-Yang-Mills source reads

\begin{equation}\protect\phantomsection\label{eq:sol-emt}{T^\mu{}_\nu = -\frac{1}{16\pi}\Bigl(\delta^\mu_\nu\,\mathcal{F}^q - 4q\,\mathrm{Tr}\bigl(\mathcal{F}^{(a)}_{\nu\lambda}\mathcal{F}^{(a)\mu\lambda}\bigr)\mathcal{F}^{q-1}\Bigr).}\end{equation}

Variation with respect to \(A^{(a)}_\mu\) gives the generalized Yang-Mills equation

\begin{equation}\protect\phantomsection\label{eq:sol-ym-eq}{\mathbf{d}\!\left({}^\star\mathbf{F}^{(a)}\mathcal{F}^{q-1}\right) + \varsigma^{-1}C^{(a)}_{(b)(c)}\mathcal{F}^{q-1}\mathbf{A}^{(b)}\wedge{}^\star\mathbf{F}^{(c)}=0,}\end{equation}

which is satisfied identically by the Wu-Yang ansatz.
Here \(\mathbf{d}\) is the exterior derivative, \({}^\star\) the Hodge dual, \(\wedge\) the wedge product, and \(\mathbf{F}^{(a)}=\tfrac{1}{2}\mathcal{F}^{(a)}_{\mu\nu}dx^\mu\wedge dx^\nu\) is the 2-form field strength \citep{Mazharimousavi:2009mb}.
Inserting the ansatz into the \(rr\) component of the Einstein equations and integrating once in \(r\), we obtain the function

\begin{equation}\protect\phantomsection\label{eq:fr}{f(r) = 1 - \frac{2M}{r} + \frac{2^{\,q-1}\,Q^{2q}}{(4q-3)\,r^{4q-2}}.}\end{equation}

Here \(M\) is the integration constant identified with the ADM (Arnowitt-Deser-Misner) mass.
This is the result of \citep{Mazharimousavi:2009mb} in four dimensions and agrees with the \(D=4\), \(\Lambda=0\) limit of the general formula given in \citep{Biswas:2022qyl, Gogoi:2023ffh}.
The formula breaks down at \(q=3/4\) where \((4q-3)\to 0\).
At this isolated point the charge term must instead be replaced by \(\propto Q^{3/2}\ln(r)/r\).
We exclude this isolated point and keep \(q\ne 3/4\) throughout.

The static seed is

\begin{equation}\protect\phantomsection\label{eq:seed}{ds_0^2 = -f(r)\,dt^2 + \frac{dr^2}{f(r)} + r^2\big(d\theta^2 + \sin^2\theta\, d\phi^2\big),}\end{equation}

with \(f(r)\) given by Eq.~\ref{eq:fr}.
The spacetime is asymptotically flat, \(f\to 1\) as \(r\to\infty\) for any \(q>1/2\) with \(q\ne 3/4\) \citep{Biswas:2022qyl, Rincon:2023hvd}.
Writing \(f(r)=1-2m(r)/r\) we read off the mass function \(m(r) = M - 2^{q-2}Q^{2q}/[(4q-3)r^{4q-3}]\), displayed in full as Eq.~\ref{eq:massfn} below.
As \(r\to\infty\) this gives \(m(r)\to M\), confirming that \(M\) is the ADM mass measured by a distant observer.
At finite \(r\) the mass function deviates from \(M\) by an amount that grows as \(r\) decreases, reflecting the energy stored in the Yang-Mills field.

The parameter \(q\) organizes a one-parameter family of spacetimes.
Setting \(q=1\) in Eq.~\ref{eq:fr} gives \(f(r)=1-2M/r+Q^2/r^2\), which is the Reissner-Nordström metric, the familiar solution of Einstein gravity sourced by a Maxwell field \citep{Mazharimousavi:2009mb, Gogoi:2023ffh}.
Setting \(Q\to 0\) at fixed \(q\) sends the charge term to zero and Eq.~\ref{eq:fr} reduces to the Schwarzschild lapse \(f=1-2M/r\).
These two limits are preserved by the rotating metric derived below, with \(q=1\) recovering the Kerr-Newman geometry and \(Q\to 0\) returning the Kerr solution.
Neither limit requires \(q\) to be an integer, so the Reissner-Nordström and Schwarzschild geometries each emerge from a single point or a line of the \((Q,q)\) parameter space rather than from isolated special cases.

The weak, the strong, and the dominant energy conditions of the power-Yang-Mills stress tensor, together with the causality condition, impose restrictions on \(q\) values that were analysed in detail by Mazharimousavi and Halilsoy \citep{Mazharimousavi:2009mb}.
In four dimensions the allowed window is

\begin{equation}\protect\phantomsection\label{eq:sol-q-range}{\frac{3}{4} \le q < \frac{3}{2}.}\end{equation}

For \(q\) below the lower bound the energy conditions are violated.
In that regime the prefactor \(1/(4q-3)\) of the charge term in Eq.~\ref{eq:fr} flips sign and the term diverges more slowly than the Coulomb one near the origin.
For \(q\ge 3/2\) causality breaks down.
The regime \(3/4<q<1\) and the complementary interval \(1<q<3/2\) are both physically admissible and will be explored in the subsequent sections.

To promote Eq.~\ref{eq:seed} to a stationary, axisymmetric spacetime we apply the complexification-free Newman-Janis algorithm of Azreg-Aïnou \citep{Azreg-Ainou:2014aqa, Azreg-Ainou:2014pra}; the motivation for this choice over the standard complexification procedure was given inSec.~\ref{sec:intro}.
Instead of complexifying \(f(r)\), we introduce undetermined metric functions of both \(r\) and the spin parameter \(a\), whose explicit forms are then fixed by requiring that the final metric take the Boyer-Lindquist form.
The procedure has been applied successfully to imperfect fluid black holes, regular black holes of Bardeen and Hayward type \citep{Bardeen:1968, Hayward:2005gi}, and various nonlinear-electrodynamics solutions \citep{Azreg-Ainou:2014pra}, and it underlies most of the rotating regular black hole metrics in current use \citep{Bambi:2013ufa, Toshmatov:2014nya}.

The first step is to rewrite Eq.~\ref{eq:seed} in advanced null (Eddington-Finkelstein) coordinates \((u,r,\theta,\phi)\) via \(du = dt - dr/f(r)\).
In these coordinates the metric reads

\begin{equation}\protect\phantomsection\label{eq:sol-nja-ef}{ds_0^2 = -f(r)\,du^2 - 2\,du\,dr + r^2\,d\Omega^2,}\end{equation}

where \(d\Omega^2=d\theta^2+\sin^2\theta\,d\phi^2\).
We then write the inverse metric as \(g^{ab}=-l^a n^b - n^a l^b + m^a\bar{m}^b + \bar{m}^a m^b\) in terms of the null tetrad components, with

\begin{equation}\protect\phantomsection\label{eq:sol-null-tetrad}{l^a=\delta^a_r, \quad n^a=\delta^a_u-\tfrac{f(r)}{2}\,\delta^a_r, \quad m^a=\frac{1}{\sqrt{2}\,r}\Bigl(\delta^a_\theta+\frac{i}{\sin\theta}\,\delta^a_\phi\Bigr).}\end{equation}

The rotation is then implemented by applying the complex shift \(u\to u-ia\cos\theta\), \(r\to r+ia\cos\theta\) to the null tetrad components, while replacing \(f(r)\) and \(r^2\) by the undetermined functions \(\mathcal{A}(r,\theta,a)\) and \(\mathcal{B}(r,\theta,a)\).
This yields a transformed null tetrad

\begin{equation}\protect\phantomsection\label{eq:sol-nja-tetrad2}{l'^a=\delta^a_r, \quad n'^a=\delta^a_u-\tfrac{\mathcal{A}}{2}\,\delta^a_r, \quad m'^a=\frac{1}{\sqrt{2\mathcal{B}}}\Bigl(ia\sin\theta\,(\delta^a_u-\delta^a_r)+\delta^a_\theta+\frac{i}{\sin\theta}\,\delta^a_\phi\Bigr).}\end{equation}

Reconstructing the metric from the new tetrad and demanding that off-diagonal terms \(g_{ur}\) and \(g_{r\phi}\) vanish after the coordinate change back to Boyer-Lindquist form requires the functions \(\mathcal{A}\) and \(\mathcal{B}\) to satisfy

\begin{equation}\protect\phantomsection\label{eq:sol-nja-ab}{\mathcal{B} = r^2 + a^2\cos^2\theta \equiv \rho^2, \qquad \mathcal{A} = \frac{r^2 f(r)+a^2\cos^2\theta}{\rho^2}.}\end{equation}

Inserting these into the metric in Eddington-Finkelstein form and performing the coordinate transformation \(dt = du + (r^2+a^2)\,dr/\Delta\), \(d\phi = d\phi + a\,dr/\Delta\) then yields the Boyer-Lindquist form.
The mass function \(m(r)\) appears through \(f(r)=1-2m(r)/r\), so the rotation step replaces \(M\) by the \(r\)-dependent function \(m(r)\) given in Eq.~\ref{eq:massfn}, the same substitution that generates Kerr-Newman from Reissner-Nordström.
Because no radial complexification is performed, the final metric is unambiguous and the freedom present in the original Newman-Janis procedure is fully removed.

Starting from Eq.~\ref{eq:seed} and introducing the spin parameter \(a\), we obtain the rotating metric in Boyer-Lindquist form.
Setting \(\rho^2 = r^2 + a^2\cos^2\theta\) as in Eq.~\ref{eq:sol-nja-ab}, the mass function is

\begin{equation}\protect\phantomsection\label{eq:massfn}{m(r) = M - \frac{2^{\,q-2}\,Q^{2q}}{(4q-3)\,r^{4q-3}},}\end{equation}

and the line element reads

\begin{equation}\protect\phantomsection\label{eq:metric}{ds^2 = -\Big(1-\frac{2 m(r)\,r}{\rho^2}\Big)dt^2 - \frac{4 a\, m(r)\, r \sin^2\theta}{\rho^2}\,dt\,d\phi + \frac{\rho^2}{\Delta}\,dr^2 + \rho^2\, d\theta^2 + \frac{\Sigma \sin^2\theta}{\rho^2}\,d\phi^2,}\end{equation}

where

\begin{equation}\protect\phantomsection\label{eq:delta}{\Delta(r) = r^2 f(r) + a^2, \qquad \Sigma = (r^2+a^2)^2 - a^2\Delta\sin^2\theta.}\end{equation}

Inserting Eq.~\ref{eq:fr} into the definition of \(\Delta\) gives

\begin{equation}\protect\phantomsection\label{eq:sol-delta-explicit}{\Delta(r) = r^2 - 2Mr + a^2 + \frac{2^{\,q-1}Q^{2q}}{(4q-3)\,r^{4q-4}}.}\end{equation}

For \(q=1\) this is the Kerr-Newman expression \(\Delta = r^2-2Mr+a^2+Q^2\), and for \(Q\to 0\) it returns the Kerr function \(\Delta=r^2-2Mr+a^2\).
When \(a\to 0\) the line element Eq.~\ref{eq:metric} reduces to the static seed Eq.~\ref{eq:seed}.
We stress that, as for any rotating metric that the Newman-Janis algorithm generates from a non-vacuum seed, the solution Eq.~\ref{eq:metric} did not solve the original EPYM field equations exactly.
We adopt it in the phenomenological spirit common to this literature, as a well-defined stationary axisymmetric geometry that carries the imprint of the power-Yang-Mills charge and reduces to the established static solution in the non-rotating limit \citep{Azreg-Ainou:2014pra, Biswas:2022qyl}.

In the orthonormal frame of Carter the Einstein tensor of Eq.~\ref{eq:metric} is diagonal and describes an anisotropic fluid with energy density \(\varepsilon\), radial pressure \(p_r\), and tangential pressure \(p_\perp\),
\begin{equation}\protect\phantomsection\label{eq:sol-eff-source}{8\pi\varepsilon = -8\pi p_r = \frac{2^{\,q-1}Q^{2q}\,r^{4-4q}}{\rho^4}, \qquad 8\pi p_\perp = \frac{2^{\,q-1}Q^{2q}\,r^{2-4q}}{\rho^4}\Big[(2q-1)\,r^2 + 2(q-1)\,a^2\cos^2\theta\Big].}\end{equation}
At \(a=0\) this is the power-Yang-Mills stress tensor Eq.~\ref{eq:sol-emt} on the Wu-Yang configuration, and at \(q=1\) it is the field of the magnetic Kerr-Newman solution, which solves the Yang-Mills equations when embedded along one generator of \(SO(3)\), so at the Maxwell point Eq.~\ref{eq:metric} is an exact Einstein-Yang-Mills solution.
For \(q\neq1\) the null, weak, and strong energy conditions hold everywhere outside the event horizon and the dominant one holds if and only if \(q\le1\), exactly as for the static seed, but we know of no power-Yang-Mills configuration that generates Eq.~\ref{eq:sol-eff-source}, as is generally the case for rotating metrics built from nonlinear-field seeds \citep{Toshmatov:2017zpr}.

\begin{figure}
\centering
\pandocbounded{\includegraphics[keepaspectratio,alt={Horizon structure of the rotating EPYM black hole, outer (r\_+, solid) and inner (r\_-, dashed) horizons against the spin a. Left: fixed charge Q=0.4\textbackslash,M\^{}\{2-1/q\} for several powers q in the admitted window. Right: fixed power q=1 for several charges Q. Each pair of curves merges at the extremal spin, beyond which no horizon exists.}]{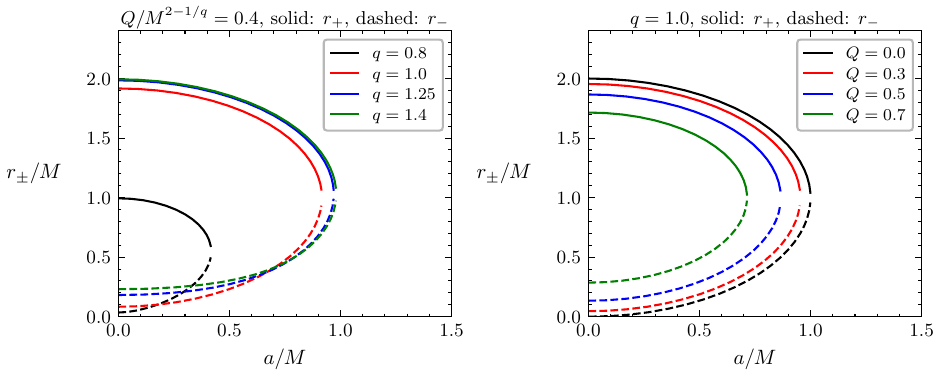}}
\caption{Horizon structure of the rotating EPYM black hole, outer (\(r_+\), solid) and inner (\(r_-\), dashed) horizons against the spin \(a\). Left: fixed charge \(Q=0.4\,M^{2-1/q}\) for several powers \(q\) in the admitted window. Right: fixed power \(q=1\) for several charges \(Q\). Each pair of curves merges at the extremal spin, beyond which no horizon exists.}\label{fig:horizon}
\end{figure}

The horizons are the positive real solutions of \(\Delta(r)=0\).
For a range of \((a,Q,q)\) the equation has two roots, the event horizon \(r_+\) outside and the Cauchy horizon \(r_-\) inside, which merge at extremality.
At fixed \(q\) and \(Q\) the extremal spin \(a_{\rm ext}\) is the largest value of \(a\) for which \(\Delta=0\) has a double root.
Beyond it no horizon exists and the solution describes a naked singularity.
As the charge \(Q\) increases at fixed \(a\) and \(q\), the event horizon \(r_+\) shrinks and the Cauchy horizon \(r_-\) grows, following the same qualitative trend as in Reissner-Nordström \citep{Rincon:2023hvd, Gogoi:2023ffh}.
The role of \(q\) is more subtle.
The charge term in \(\Delta\) stays positive for every admitted \(q\), but near the horizon it weakens monotonically with \(q\) at fixed \(Q/M^{2-1/q}\), both through the prefactor \(2^{q-1}/(4q-3)\) and through the faster falloff \(r^{4-4q}\), so \(r_+\) rises monotonically toward the Kerr value, while staying below it, as \(q\) sweeps the window.
Setting \(Q\to0\) removes the charge term from \(\Delta\) and the standard Kerr horizon \(r_\pm=M\pm\sqrt{M^2-a^2}\) is recovered in agreement with the Schwarzschild limit discussed earlier.
The behaviour of both horizons across the parameter space is shown inFig.~\ref{fig:horizon}.

The three qualitatively distinct configurations (a two-horizon black hole, an extremal black hole with one degenerate horizon, and a horizonless naked singularity) partition the \((Q,q)\) parameter plane at fixed spin into distinct phases.
The boundary between the black hole and naked singularity phases is the extremal curve \(a_{\rm ext}(Q,q)\), defined by the simultaneous conditions \(\Delta(r_{\rm ext})=0\) and \(\partial_r\Delta(r_{\rm ext})=0\).
Since \(\partial_r\Delta\) is independent of \(a\), we can find \(r_{\rm ext}(Q,q)\) by minimizing \(\Delta\) with respect to \(r\) at fixed \((Q,q)\), and then read off

\begin{equation}a_{\rm ext}^2 = 2Mr_{\rm ext} - r_{\rm ext}^2 - 2^{q-1}Q^{2q}/[(4q-3)r_{\rm ext}^{4q-4}].\quad \end{equation}

\begin{figure}
\centering
\pandocbounded{\includegraphics[keepaspectratio,alt={Phase diagram of the rotating EPYM black hole in the (Q/M\^{}\{2-1/q\},q) plane at fixed spin a=0.7\textbackslash,M. The plain blue region marks parameter pairs (Q,q) for which a two-horizon black hole exists and the hatched red region those with no horizon (naked singularity), as named in the legend. The solid contour is the extremal boundary a\_\{\textbackslash rm ext\}(Q,q)=0.7\textbackslash,M, where the outer and inner horizons coalesce into one degenerate horizon. The dotted line marks the Maxwell point q=1.}]{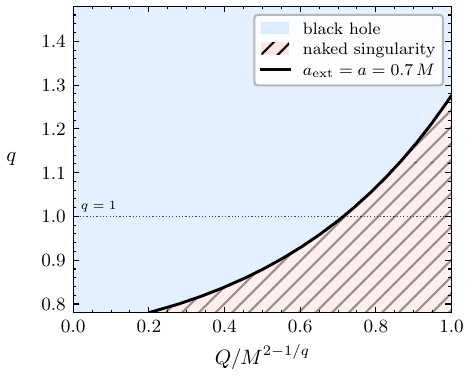}}
\caption{Phase diagram of the rotating EPYM black hole in the \((Q/M^{2-1/q},q)\) plane at fixed spin \(a=0.7\,M\). The plain blue region marks parameter pairs \((Q,q)\) for which a two-horizon black hole exists and the hatched red region those with no horizon (naked singularity), as named in the legend. The solid contour is the extremal boundary \(a_{\rm ext}(Q,q)=0.7\,M\), where the outer and inner horizons coalesce into one degenerate horizon. The dotted line marks the Maxwell point \(q=1\).}\label{fig:phase_diagram}
\end{figure}

For \(a < a_{\rm ext}\) the black hole carries two horizons and at \(a = a_{\rm ext}\) they coalesce into one degenerate horizon and for \(a > a_{\rm ext}\) the singularity is naked.
These regions are shown in Fig.~\ref{fig:phase_diagram}, which maps the \((Q/M^{2-1/q},q)\) plane at fixed spin \(a=0.7\,M\).
Parameter pairs left of the extremal contour (blue region) correspond to a two-horizon black hole.
Those right of it (hatched red region) admit no horizon.
The Maxwell point \(q=1\) is marked.

\begin{figure}
\centering
\pandocbounded{\includegraphics[keepaspectratio,alt={Ergoregion of the rotating EPYM black hole in the poloidal (x,z) plane, for (a,Q,q)=(0.7,0.4,1), (0.9,0.4,1), and (0.9,0.4,1.25) from left to right, with a in units of M and Q in units of M\^{}\{2-1/q\}. In each panel the static limit surface (outer curve, \textbackslash Delta=a\^{}2\textbackslash sin\^{}2\textbackslash theta) encloses the event horizon (inner circle, r\_+) and meets it at the poles, and the shaded gap is the ergoregion in which no static observer can exist. Raising the spin from the first panel to the second widens the equatorial cross-section, while raising the power to q=1.25 in the third narrows it again.}]{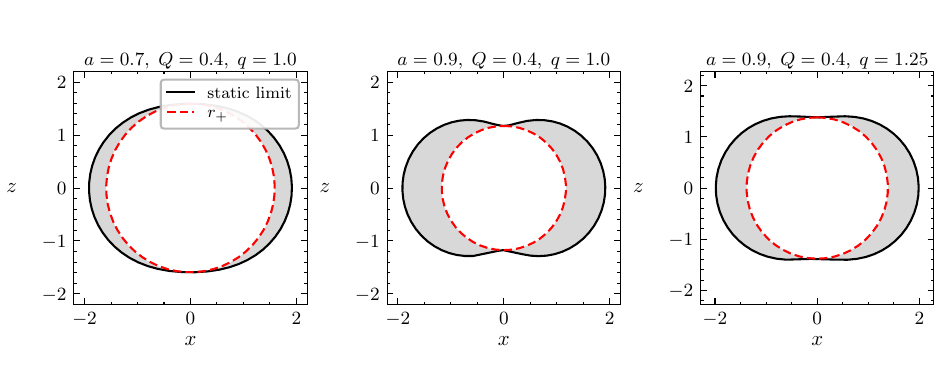}}
\caption{Ergoregion of the rotating EPYM black hole in the poloidal \((x,z)\) plane, for \((a,Q,q)=(0.7,0.4,1)\), \((0.9,0.4,1)\), and \((0.9,0.4,1.25)\) from left to right, with \(a\) in units of \(M\) and \(Q\) in units of \(M^{2-1/q}\). In each panel the static limit surface (outer curve, \(\Delta=a^2\sin^2\theta\)) encloses the event horizon (inner circle, \(r_+\)) and meets it at the poles, and the shaded gap is the ergoregion in which no static observer can exist. Raising the spin from the first panel to the second widens the equatorial cross-section, while raising the power to \(q=1.25\) in the third narrows it again.}\label{fig:ergosphere}
\end{figure}

In addition to the event horizon, the rotating metric Eq.~\ref{eq:metric} admits a static limit surface defined by the vanishing of \(g_{tt}\).
From Eq.~\ref{eq:metric} this corresponds to

\begin{equation}\protect\phantomsection\label{eq:sol-static-limit}{\rho^2 - 2m(r)\,r = 0,}\end{equation}

which is equivalent to \(\Delta(r) = a^2\sin^2\theta\).
At the poles (\(\theta=0,\pi\)) the static limit surface coincides with \(r_+\), but away from the poles it lies outside the event horizon.
The region between the static limit surface and \(r_+\) is the ergoregion, where \(g_{tt}>0\) and no static observer can exist.
All matter and radiation must corotate with the black hole, while the Penrose process can extract rotational energy from this region \citep{Penrose:1969pc}.
The ergoregion is present for any \(a>0\) and its radial extent depends on both \(Q\) and \(q\) through the mass function \(m(r)\).
For \(a=0\) the static limit surface collapses onto the horizon (at \(q=1\) recovering the Reissner-Nordström geometry) with no ergoregion as expected.
The static limit surface and the enclosed ergoregion are displayed in Fig.~\ref{fig:ergosphere}.

The spacetime described by Eq.~\ref{eq:metric} contains a curvature singularity at \(r=0\).
For the static seed the Ricci scalar diverges as

\begin{equation}\protect\phantomsection\label{eq:sol-ricci}{\mathcal{R} \sim \frac{(2Q^2)^q\,(2-2q)}{r^{4q}}, \qquad r\to 0,}\end{equation}

which diverges for all \(q\) in the range Eq.~\ref{eq:sol-q-range} except at the Maxwell point \(q=1\), where the Ricci scalar vanishes as it must for the Reissner-Nordström solution.
At \(q=1\) the singularity shows up instead in the Kretschmann scalar \(R_{\mu\nu\alpha\beta}R^{\mu\nu\alpha\beta}\), which grows as \(r^{-8q}\) across the whole window.
The rotating solution inherits the same ring nature of the central singularity of Kerr-Newman geometries\citep{Newman:1965my} .

At large \(r\) the metric Eq.~\ref{eq:metric} is asymptotically flat.
The charge term in Eq.~\ref{eq:fr} falls off as \(r^{2-4q}\), which decays faster than \(1/r\) for \(q>3/4\), so \(f(r)\to 1\) and the geometry approaches Minkowski space at spatial infinity.
The ADM mass is \(M\) and the metric is Kerr-like at leading order in \(1/r\), making it suitable for astrophysical applications in which the observer is far from the source.

\section{Shadow and photon region}\label{sec:shadow}

The shadow is fixed by the unstable photon orbits that the metric Eq.~\ref{eq:metric} supports.
We obtain the null geodesics from the Hamilton-Jacobi equation, which separates for this Kerr-like geometry \citep{Shaikh:2019fpu}.
Writing the Jacobi action as

\begin{equation}\protect\phantomsection\label{eq:jacobi}{\mathcal{S}=-Et+L\phi+S_r(r)+S_\theta(\theta),}\end{equation}

where \(S_r\) and \(S_\theta\) are the separated radial and angular parts, with \(E\) and \(L\) the conserved energy and axial angular momentum, the separation introduces the Carter constant \(\mathcal{K}\) and yields the radial equation \(\rho^4\dot{r}^2=\mathcal{P}(r)\) with

\begin{equation}\protect\phantomsection\label{eq:radialpot}{\mathcal{P}(r) = \big[(r^2+a^2)E - aL\big]^2 - \Delta(r)\big[\mathcal{K} + (L-aE)^2\big].}\end{equation}

We derive these equations starting from the Hamilton-Jacobi equation \(\partial\mathcal{S}/\partial\lambda + H = 0\), where the Hamiltonian is \(H = \frac{1}{2}g^{\mu\nu}p_\mu p_\nu\) and \(\lambda\) is the affine parameter \citep{Shaikh:2019fpu}.
Because the metric Eq.~\ref{eq:metric} is independent of \(t\) and \(\phi\), the two first integrals \(E=-p_t\) and \(L=p_\phi\) arise immediately from the Killing vectors \(\partial_t\) and \(\partial_\phi\) of the geometry.
Setting the particle mass to zero for photons and substituting the separability ansatz into the geodesic equations yields, together with the radial equation Eq.~\ref{eq:radialpot}, the angular equation

\begin{equation}\protect\phantomsection\label{eq:sh-angular}{\rho^4\dot\theta^2 = \Theta(\theta),}\end{equation}

with

\begin{equation}\protect\phantomsection\label{eq:sh-thetapot}{\Theta(\theta) = \mathcal{K} + a^2 E^2\cos^2\theta - L^2\cot^2\theta.}\end{equation}

The quantities \(\mathcal{P}(r)\) and \(\Theta(\theta)\) must both be non-negative along any physical photon trajectory.
The Carter constant \(\mathcal{K}\) is the separation constant that makes the geodesic system fully integrable \citep{Carter:1968rr} and it has no Newtonian analogue and encodes the latitudinal motion.
For equatorial geodesics we have \(\theta=\pi/2\) and \(\Theta(\pi/2)=\mathcal{K}\) and equatorial confinement requires \(\dot\theta=0\), so \(\Theta(\pi/2)=0\) and the Carter constant vanishes \(\mathcal{K}=0\).

A convenient way to classify photon trajectories is to rewrite the radial equation Eq.~\ref{eq:radialpot} as \(\rho^4\dot{r}^2 + V_{\rm eff}(r) = 0\), where the conserved energy is scaled out by the affine reparameterization \(\lambda \to E\lambda\).
Defining the reduced radial potential by factoring out \(E^2\) we obtain

\begin{equation}\protect\phantomsection\label{eq:sh-veff}{V_{\rm eff}(r) = -\frac{\mathcal{P}(r)}{E^2} = -\Big[(r^2+a^2) - a\xi\Big]^2 + \Delta(r)\Big[\eta + (\xi-a)^2\Big],}\end{equation}

where \(\xi = L/E\) and \(\eta = \mathcal{K}/E^2\) are Bardeen\textquotesingle s impact parameters, carrying units of \(M\) and \(M^2\) in geometrized units \citep{Bardeen:1972fi}.
Photons with \(V_{\rm eff}<0\) at a given \(r\) are able to move radially at that point; photons at a maximum of \(V_{\rm eff}\) are momentarily trapped.
Unstable circular photon orbits sit at radii where \(\mathcal{P}=0\) and \(\mathcal{P}'=0\) simultaneously, with \(\mathcal{P}''>0\) confirming instability \citep{Gralla:2019xty}.
These three conditions define the photon sphere in the spherically symmetric limit and its rotating generalisation, the photon region, in the axisymmetric case \citep{Grenzebach:2014fha}.
Solving the first two conditions for \(\xi\) and \(\eta\) eliminates the orbital radius in favour of the impact parameters.

Spherical photon orbits sit at the radius where \(\mathcal{P}=\mathcal{P}'=0\).
Solving these two conditions gives the two impact parameters \(\xi=L/E\) and \(\eta=\mathcal{K}/E^2\) as functions of the orbit radius \(r\),

\begin{equation}\protect\phantomsection\label{eq:xi}{\xi(r) = \frac{(r^2+a^2)\,\Delta'(r) - 4r\,\Delta(r)}{a\,\Delta'(r)},}\end{equation}

\begin{equation}\protect\phantomsection\label{eq:eta}{\eta(r) = \frac{r^2\big[\,8r\,\Delta(r)\Delta'(r) - 16\,\Delta(r)\big(\Delta(r)-a^2\big) - r^2\,\Delta'(r)^2\,\big]}{a^2\,\Delta'(r)^2}.}\end{equation}

For a rotating black hole these orbits do not lie on a single sphere.
They fill a photon region of finite radial width, bounded by the radii \(r_{p,\min}\) and \(r_{p,\max}\) that solve \(\eta(r)=0\).
The shadow edge is the projection of this region onto the observer\textquotesingle s celestial plane.

To derive Eq.~\ref{eq:xi} and Eq.~\ref{eq:eta} explicitly we substitute the conditions \(\mathcal{P}(r_p)=0\) and \(\mathcal{P}'(r_p)=0\) for the rotating EPYM metric.
The first condition gives

\begin{equation}[(r^2+a^2)-a\xi]^2 = \Delta\,[\eta+(\xi-a)^2].\quad \end{equation}

Differentiating \(\mathcal{P}\) with respect to \(r\) and setting the result to zero yields a second linear relation between \(\xi\) and \(\eta\) involving \(\Delta'\).
Eliminating \(\eta\) between the two equations produces Eq.~\ref{eq:xi}.
Substituting back then produces Eq.~\ref{eq:eta}.
The derivation follows exactly the same algebraic route used for the Kerr-Newman case \citep{Shaikh:2019fpu}, because the metric Eq.~\ref{eq:metric} has the same \(\Delta\)-structure: only the explicit form of \(\Delta(r)\) through Eq.~\ref{eq:fr} changes.
Note that the denominator \(\Delta'(r)^2\) in both expressions can vanish at the degenerate horizon of the extremal configuration \(\Delta=\Delta'=0\), where the two horizons merge.
The boundary radii \(r_{p,\min}\) and \(r_{p,\max}\) of the photon region merge instead when \(\eta(r)=0\) develops a double root, a different condition.

In the rotating case the static photon sphere widens into a photon region with a continuous family of spherical photon orbits parameterised by their radius \(r_p\in[r_{p,\min},r_{p,\max}]\) \citep{Grenzebach:2014fha}.
The boundary radii \(r_{p,\min}\) and \(r_{p,\max}\) are the pair of positive real roots of Eq.~\ref{eq:eta}.
Prograde orbits (co-rotating with the black hole) sit at smaller \(r_p\) and retrograde orbits sit at larger \(r_p\); for \(Q\to 0\), \(a\to 0\) the two roots merge to a single photon sphere at \(r_p=3M\).
The shadow silhouette is swept out by running \(r_p\) from \(r_{p,\min}\) to \(r_{p,\max}\) through Eq.~\ref{eq:xi} and Eq.~\ref{eq:eta} and then mapping to celestial coordinates.
Both the photon region width and the resulting \(D\)-shape of the shadow grow with increasing spin \(a\).

An observer at inclination \(\theta_o\) sees the boundary through the celestial coordinates

\begin{equation}\protect\phantomsection\label{eq:celestial}{\alpha = -\frac{\xi}{\sin\theta_o}, \qquad \beta = \pm\sqrt{\,\eta + a^2\cos^2\theta_o - \xi^2\cot^2\theta_o\,},}\end{equation}

with \(r\) running over the photon region as a parameter.
For an equatorial observer \(\theta_o=\pi/2\) these reduce to \(\alpha=-\xi\) and \(\beta=\pm\sqrt{\eta}\).
Obviously, when \(Q\to 0\) the curve \((\alpha,\beta)\) collapses onto the Kerr shadow, and a further \(a\to 0\) gives the circular Schwarzschild shadow of radius \(3\sqrt{3}\,M\).
The power-Yang-Mills charge deforms this baseline.
The charge is repulsive for every \(q\) in the admitted window Eq.~\ref{eq:sol-q-range}, contracting the shadow as \(Q\) grows \citep{Biswas:2022qyl, Zubair:2023cep}.

In the static (\(a=0\)) limit the photon region collapses to a single radius \(r_c\) satisfying \(2f(r_c)=r_c f'(r_c)\), and the shadow becomes a circle.
The shadow radius for a distant observer then follows from the critical impact parameter

\begin{equation}\protect\phantomsection\label{eq:sh-rph-static}{R_s^{\rm static} = \frac{r_c}{\sqrt{f(r_c)}}.}\end{equation}

This formula is the standard result \citep{Gralla:2019xty, Jusufi:2019ltj}, where \(f(r)\) is given by Eq.~\ref{eq:fr} with \(a=0\).
Inserting Eq.~\ref{eq:fr} into \(2f=rf'\) yields an algebraic equation for \(r_c\) that must be solved numerically for general \(q\).
For \(q=1\) it reduces to the Reissner-Nordström photon sphere radius.

We verify the limits explicitly.
For \(Q\to 0\) the mass function reduces to \(m(r)=M\), so \(\Delta\to r^2-2Mr+a^2\) and Eq.~\ref{eq:xi} and Eq.~\ref{eq:eta} reduce to the standard Kerr expressions \citep{Shaikh:2019fpu}

\begin{equation}\protect\phantomsection\label{eq:sh-kerr-limits}{\xi_{\rm Kerr} = \frac{(r^2+a^2)(r-M) - 2r\Delta_{\rm Kerr}}{a(r-M)}, \qquad \eta_{\rm Kerr} = \frac{r^3\big[4a^2 M - r(r-3M)^2\big]}{a^2(r-M)^2}.}\end{equation}

A further \(a\to 0\) collapses both roots of \(\eta=0\) to \(r_c=3M\) and the shadow becomes a perfect circle of radius \(3\sqrt{3}M\) in the celestial plane.
We set \(q=1\) as a second cross-check and recover the Kerr-Newman shadow, for which an extensive literature exists \citep{Vagnozzi:2022moj, Kumar:2020owy, Tsukamoto:2017fxq, Amir:2016cen}.
The shadow has been computed for a wide range of rotating and deformed metrics, from braneworld and overspinning solutions to regular black holes, and proposed as a test of the metric beyond Kerr \citep{Cunha:2018acu, Johannsen:2010ru, Bambi:2008jg, Amarilla:2011fx, Abdujabbarov:2016hnw}.
These two reductions confirm that Eq.~\ref{eq:xi} and Eq.~\ref{eq:eta} as derived for the EPYM metric are consistent with known results.

We describe the boundary by two coordinate-independent observables \citep{Kumar:2020owy}.
The first is the area inside the silhouette,

\begin{equation}\protect\phantomsection\label{eq:area}{A = 2\int_{r_{p,\min}}^{r_{p,\max}} \beta(r)\,\frac{d\alpha(r)}{dr}\,dr,}\end{equation}

and the corresponding areal radius reads

\begin{equation}\protect\phantomsection\label{eq:areal-radius}{R_s = \sqrt{\frac{A}{\pi}}.}\end{equation}

The second is the oblateness \(D\)

\begin{equation}\protect\phantomsection\label{eq:oblateness}{D=\frac{\Delta\alpha}{\Delta\beta},}\end{equation}

where \(\Delta\alpha=\alpha_{\max}-\alpha_{\min}\) and \(\Delta\beta=\beta_{\max}-\beta_{\min}\) are the horizontal and vertical extents of the shadow boundary.
The ratio equals one for a circular shadow and falls below one as the spin flattens the left edge.
These two numbers carry the dependence on \((a,Q,q,\theta_o)\) and are what we confront with the data in the next section.

Beyond the areal radius \(R_s\) and the oblateness \(D\), we can characterise the shadow by a distortion parameter \(\delta_s\) that measures the fractional departure of the leftmost boundary point from the reference circle of radius \(R_s\) \citep{Hioki:2009na}.
Such coordinate-independent observables underlie the estimation of black-hole parameters from the shadow boundary \citep{Kumar:2018ple, Tsupko:2017rdo}.
Defining \(\tilde{\alpha}_l\) as the leftmost point on the distorted contour and \(\alpha_l\) as its counterpart on the reference circle, we write

\begin{equation}\protect\phantomsection\label{eq:sh-distortion}{\delta_s = \frac{|\tilde{\alpha}_l - \alpha_l|}{R_s}.}\end{equation}

\begin{figure}
\centering
\pandocbounded{\includegraphics[keepaspectratio,alt={Hioki-Maeda distortion parameter \textbackslash delta\_s Eq.~ of the rotating EPYM shadow for an equatorial observer, computed with the reference circle of areal radius R\_s anchored at the rightmost point of the silhouette. Left: \textbackslash delta\_s against spin a for several powers q at fixed Q=0.3\textbackslash,M\^{}\{2-1/q\}, with the Kerr curve (Q=0) dashed. Right: \textbackslash delta\_s against spin a for several charges Q at fixed q=1, with the Kerr curve dashed. Curves terminate where the configuration becomes horizonless.}]{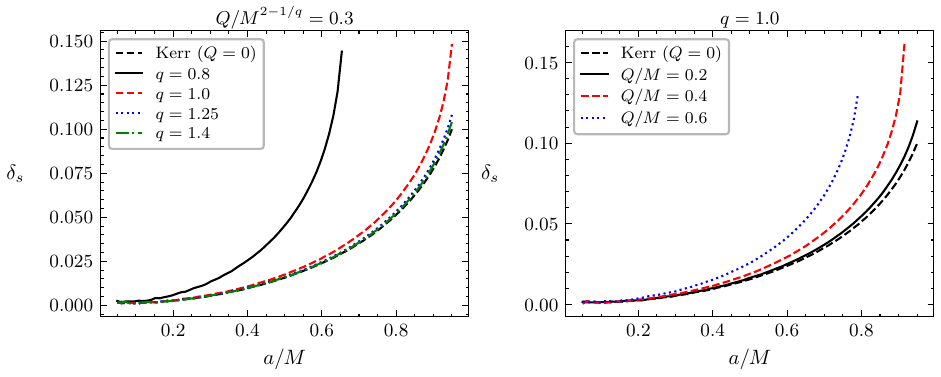}}
\caption{Hioki-Maeda distortion parameter \(\delta_s\) Eq.~\ref{eq:sh-distortion} of the rotating EPYM shadow for an equatorial observer, computed with the reference circle of areal radius \(R_s\) anchored at the rightmost point of the silhouette. Left: \(\delta_s\) against spin \(a\) for several powers \(q\) at fixed \(Q=0.3\,M^{2-1/q}\), with the Kerr curve (\(Q=0\)) dashed. Right: \(\delta_s\) against spin \(a\) for several charges \(Q\) at fixed \(q=1\), with the Kerr curve dashed. Curves terminate where the configuration becomes horizonless.}\label{fig:shadow_distortion}
\end{figure}

For a circular shadow \(\delta_s=0\).
The quantity \(\delta_s\) increases with spin and, at fixed spin, with the charge \(Q\) for every \(q\) in the window, the increase being steeper at small \(q\) and weaker as \(q\) approaches \(3/2\).
An alternative and simpler observable, free of any coordinate choice, is the linear shadow radius \(R_s\) together with the horizontal extent \(\Delta\alpha\) (defined above), both of which can be read directly from the parametric curve without fitting a reference circle.
These trends are confirmed numerically inFig.~\ref{fig:shadow_distortion}, the distortion grows with the spin at every admitted \(q\), fastest near the lower edge of the window, and at fixed \(q\) it grows with the charge.
We use \(R_s\) and \(D\) as the primary observables confronted with EHT data inSec.~\ref{sec:eht}, because they are directly related to the observed angular size and circularity deviation reported by the collaboration.

The qualitative behaviour of the shadow with the three parameters \((a,Q,q)\) can be understood directly from the structure of \(\Delta(r)\) in Eq.~\ref{eq:fr}.
Increasing the spin \(a\) drags the prograde photon orbit inward and pushes the retrograde one outward, resulting in a larger horizontal width and a more pronounced \(D\)-shape.
The dependence on the charge \(Q\) and the power parameter \(q\) is more subtle, but it does not reverse across the Maxwell point \(q=1\).
In the admitted window Eq.~\ref{eq:sol-q-range} the charge term in \(f(r)\) carries the positive prefactor \(2^{q-1}/(4q-3)\), so it is repulsive for every \(q\): increasing \(Q\) pushes the unstable photon orbit inward and contracts the shadow, and \(R_s\) decreases monotonically with \(Q\) at all \(q\).
What changes across \(q=1\) is not the sign but the rate.
For \(q\) close to the lower bound \(3/4\) the prefactor \(1/(4q-3)\) is large and the charge term falls off slowly, so the shadow contracts sharply with \(Q\); for larger \(q\) the prefactor shrinks and the falloff \(r^{-(4q-2)}\) is faster, so the contraction is mild.
This is consistent with the effective potential analysis: the peak of \(V_{\rm eff}\) moves toward the horizon as \(Q\) increases for every \(q\), only by a larger amount at small \(q\) \citep{Zubair:2023cep}.
Obviously, the threshold \(q=3/4\) bounds the window from below, and since the metric function is singular there, we keep \(q\) strictly above it.

\begin{figure}
\centering
\pandocbounded{\includegraphics[keepaspectratio,alt={Shadow boundary (\textbackslash alpha,\textbackslash beta) of the rotating EPYM black hole for an equatorial observer at fixed spin a=0.7\textbackslash,M. Left: fixed Q=0.3\textbackslash,M\^{}\{2-1/q\} and several powers q spanning the Maxwell point, with the Kerr reference (Q=0) dashed. Right: fixed q=1 and several charges Q, with the Kerr curve dashed. The charge contracts the silhouette and the contraction rate orders by q.}]{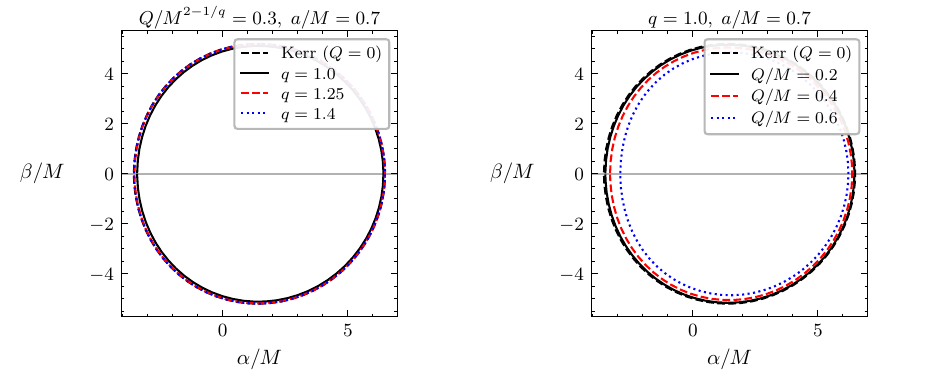}}
\caption{Shadow boundary \((\alpha,\beta)\) of the rotating EPYM black hole for an equatorial observer at fixed spin \(a=0.7\,M\). Left: fixed \(Q=0.3\,M^{2-1/q}\) and several powers \(q\) spanning the Maxwell point, with the Kerr reference (\(Q=0\)) dashed. Right: fixed \(q=1\) and several charges \(Q\), with the Kerr curve dashed. The charge contracts the silhouette and the contraction rate orders by \(q\).}\label{fig:shadow_contours}
\end{figure}

\begin{figure}
\centering
\pandocbounded{\includegraphics[keepaspectratio,alt={Areal shadow radius R\_s=\textbackslash sqrt\{A/\textbackslash pi\} of the rotating EPYM black hole against the spin a for an equatorial observer. Left: fixed Q=0.3\textbackslash,M\^{}\{2-1/q\} and several powers q, with the Kerr curve (Q=0) dashed. Right: fixed q=1 and several charges Q, with the Kerr curve dashed. Curves terminate where the configuration becomes horizonless.}]{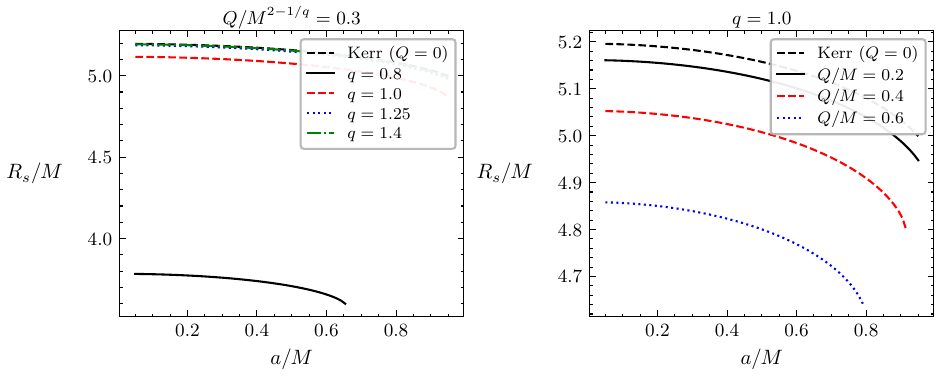}}
\caption{Areal shadow radius \(R_s=\sqrt{A/\pi}\) of the rotating EPYM black hole against the spin \(a\) for an equatorial observer. Left: fixed \(Q=0.3\,M^{2-1/q}\) and several powers \(q\), with the Kerr curve (\(Q=0\)) dashed. Right: fixed \(q=1\) and several charges \(Q\), with the Kerr curve dashed. Curves terminate where the configuration becomes horizonless.}\label{fig:shadow_radius}
\end{figure}

The spin enters differently: increasing \(a\) always reduces the prograde photon sphere radius and always increases the shadow width through the growing separation between the prograde and retrograde boundary radii.
The two figures below capture these trends numerically for a representative set of parameters.
The shadow boundary and the areal radius are shown inFig.~\ref{fig:shadow_contours} andFig.~\ref{fig:shadow_radius} respectively.

For a distant observer the shadow also fixes the limiting absorption cross section of the black hole \citep{Wei:2013kza}.
At high frequencies the cross section for absorption oscillates about a limiting constant \(\sigma_{\rm lim}\) set by the photon region, and since the shadow is the outer boundary of that region its value approaches the geometric area of the silhouette \citep{Decanini:2011xi},

\begin{equation}\protect\phantomsection\label{eq:sh-sigma-lim}{\sigma_{\rm lim} \approx \pi R_s^2,}\end{equation}

with \(R_s\) the areal shadow radius of Eq.~\ref{eq:area}.
The energy radiated per unit frequency and unit time by the evaporating black hole then takes the form \citep{Wei:2013kza}

\begin{equation}\protect\phantomsection\label{eq:sh-emission}{\frac{d^2\mathcal{E}(\omega)}{d\omega\,dt} = \frac{2\pi^3 R_s^2\,\omega^3}{e^{\omega/T_H}-1},}\end{equation}

where \(\omega\) denotes the emission frequency and \(T_H\) the Hawking temperature.
We read \(T_H\) off the horizon surface gravity \(\kappa\),

\begin{equation}\protect\phantomsection\label{eq:sh-hawking}{T_H = \frac{\kappa}{2\pi}, \qquad \kappa = \frac{\Delta'(r_+)}{2\,(r_+^2+a^2)},}\end{equation}

which reduces to \(\kappa=\tfrac{1}{2}f'(r_+)\) in absence of rotation.
The cross section enters Eq.~\ref{eq:sh-emission} only through \(R_s\), so the whole charge and spin dependence of the emission is carried by \(R_s\) and \(T_H\).

For \(a\neq0\) the expression Eq.~\ref{eq:sh-emission} neglects the greybody factors, azimuthal modes and superradiant emission of a rotating hole, so the curves at \(a\neq0\) are only a high-frequency estimate.

\begin{figure}
\centering
\pandocbounded{\includegraphics[keepaspectratio,alt={Energy emission rate d\^{}2\textbackslash mathcal\{E\}/d\textbackslash omega\textbackslash,dt of the rotating EPYM black hole against the frequency \textbackslash omega at fixed spin a=0.7\textbackslash,M. Left: fixed Q=0.3\textbackslash,M\^{}\{2-1/q\} and several powers q spanning the Maxwell point, with the Kerr curve (Q=0) dashed. Right: fixed q=1 and several charges Q, with the Kerr curve dashed. The peak height decreases with Q and increases with q.}]{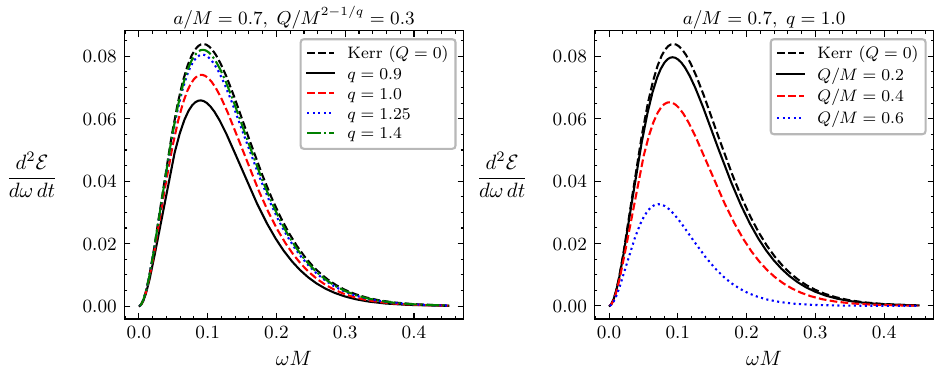}}
\caption{Energy emission rate \(d^2\mathcal{E}/d\omega\,dt\) of the rotating EPYM black hole against the frequency \(\omega\) at fixed spin \(a=0.7\,M\). Left: fixed \(Q=0.3\,M^{2-1/q}\) and several powers \(q\) spanning the Maxwell point, with the Kerr curve (\(Q=0\)) dashed. Right: fixed \(q=1\) and several charges \(Q\), with the Kerr curve dashed. The peak height decreases with \(Q\) and increases with \(q\).}\label{fig:energy_emission}
\end{figure}

The emission rate is shown inFig.~\ref{fig:energy_emission}.
Each curve is a Planck type profile with a single peak near \(\omega\approx3T_H\). Increasing the charge \(Q\) lowers both \(R_s\) and \(T_H\), so the peak drops and shifts slightly to lower frequency, which corresponds to a slower evaporation and a longer lifetime of the black hole. The power parameter \(q\) acts in the opposite direction. For larger \(q\) the shadow radius and the temperature both grow, the peak rises, and the evaporation speeds up.

\section{Constraints from EHT observations}\label{sec:eht}

Because the photon orbits of a rotating black hole are not planar, the shadow silhouette is no longer a perfect circle.
A characterisation of its size and shape that requires no choice of coordinates employs two quantities computed from the shadow boundary curve \((\alpha(r), \beta(r))\) in the observer\textquotesingle s sky plane.

We compute \(A\) by numerically integrating the shadow contour at each \((Q, q, a, \theta_o)\) and infer \(R_s\) from Eq.~\ref{eq:areal-radius}.
Note that \(D\) departs from unity only when the inclination is appreciable, for example at \(\theta_o = 17^\circ\) the departure is small, while at \(50^\circ\) it becomes observationally relevant.

The oblateness itself turns out to carry little constraining power for this family.
Across the parameter space admitted below, the predicted \(D\) stays close to unity, \(1-D\lesssim0.011\) at \(\theta_o=17^\circ\) and \(1-D\lesssim0.06\) at \(\theta_o=50^\circ\) for spins up to \(0.9\,M\), comfortably inside the \(\lesssim10\%\) deviation from circularity reported for the M87\(^*\) ring \citep{EventHorizonTelescope:2019dse}.
The oblateness therefore acts as a consistency check rather than an extra constraint, and the quantitative bounds below rest on the angular diameter \(\theta_d\) and the deviation parameter \(\delta\).

The shadow size translates into an angular diameter once the mass and the distance of a source are fixed.
For a black hole of areal shadow radius \(R_s\) at distance \(d\), the image subtends an angular diameter

\begin{equation}\protect\phantomsection\label{eq:angdia}{\theta_d = \frac{2 R_s\, M}{d},}\end{equation}

where \(d\) denotes the source distance, \(R_s\) is measured in units of \(M\) through Eq.~\ref{eq:area} and the product \(R_s M\) restores the length scale \citep{Kumar:2020owy, Vagnozzi:2022moj}.
It is convenient to quote the measured diameter also in units of the angular gravitational radius \(\theta_g = GM/(c^2 d)\).
This defines the dimensionless shadow diameter \(d_\mathrm{sh} = \theta_d/\theta_g = 2R_s\), with the Schwarzschild value \(d_\mathrm{sh} = 6\sqrt{3}\).
Given \((a,q,\theta_o)\), equation Eq.~\ref{eq:angdia} makes \(\theta_d\) a function of the charge \(Q\), so the measured angular size traces out an allowed band in the parameter space.
We compare with the two EHT targets in turn.

For M87\(^*\) we take \(M = 6.5 \times 10^9\,M_\odot\) and \(d = 16.8\,\mathrm{Mpc}\), and we view the source nearly face-on at \(\theta_o = 17^\circ\) \citep{EventHorizonTelescope:2019dse, EventHorizonTelescope:2019ggy}.
The EHT measured an emission ring of angular diameter \(\theta_d = (42 \pm 3)\,\mu\mathrm{as}\) at the \(1\sigma\) level, which gives \(d_\mathrm{sh}^{\mathrm{M87}^*} = (11.0 \pm 1.5)\,M\) once the uncertainties in the mass and the distance are combined with the imaging error \citep{Kumar:2020owy, Vagnozzi:2022moj}.
The low inclination keeps the shadow nearly circular, so the constraint on \(R_s\) translates almost entirely into a constraint on the combination of \(Q\) and \(q\) that controls the shadow area rather than the shape.
We compute \(\theta_d(Q; a, q, \theta_o)\) numerically via Eq.~\ref{eq:angdia} and Eq.~\ref{eq:areal-radius} and identify the values of \(Q\) for which the curve enters or exits the measured \(1\sigma\) band.

For Sgr A\(^*\) we take \(M \simeq 4.0 \times 10^6\,M_\odot\) and \(d \simeq 8.28\,\mathrm{kpc}\) \citep{Abuter:2018uum, EventHorizonTelescope:2022wkp}, with an inclination \(\theta_o \simeq 50^\circ\) \citep{EventHorizonTelescope:2022xqj}.
The EHT measured the angular shadow diameter to be \(\theta_d = (48.7 \pm 7)\,\mu\mathrm{as}\), corresponding in geometrized units to \(d_\mathrm{sh}^{\mathrm{Sgr}\,\mathrm{A}^*} = (10.2 \pm 1.5)\,M\).

Because decades of stellar-orbit monitoring tightly constrain the Sgr A\(^*\) mass-to-distance ratio \citep{Abuter:2018uum, Do:2019txf, EventHorizonTelescope:2022xqj}, the EHT paper VI additionally quantifies the constraint through the fractional Schwarzschild deviation parameter

\begin{equation}\protect\phantomsection\label{eq:eht-delta}{\delta \equiv \frac{d_\mathrm{sh}}{d_\mathrm{sh}^\mathrm{Schw}} - 1,}\end{equation}

where \(d_\mathrm{sh} = \theta_d/\theta_g\) is the dimensionless shadow diameter introduced above and \(d_\mathrm{sh}^\mathrm{Schw} = 6\sqrt{3}\) is its Schwarzschild value \citep{EventHorizonTelescope:2022xqj}.
The EHT VI result is \(\delta = -0.08^{+0.09}_{-0.09}\), consistent with zero and thus with the Kerr prediction.
We translate this bound into a constraint on the EPYM parameters by computing \(\delta(Q; a, q)\) at fixed \(M = 1\) and comparing with the \(1\sigma\) range \(-0.17 \lesssim \delta \lesssim 0.01\).
The larger fractional uncertainty in the Sgr A\(^*\) angular diameter, relative to M87\(^*\), makes the absolute bound on \(Q\) somewhat weaker.

A key feature of the power-Yang-Mills model is that \(\partial R_s/\partial Q < 0\) for every \(q\) in the admitted window Eq.~\ref{eq:sol-q-range}: the charge term in the lapse carries the positive prefactor \(2^{q-1}/(4q-3)\), so it is repulsive at all \(q\), contracts the photon orbit, and shrinks the shadow.
The angular size \(\theta_d\) is therefore a decreasing function of \(Q\) at every \(q\), and the lower edge of the observed EHT band, \(\theta_d^\mathrm{min}\), sets a direct upper bound on \(Q\).
What depends on \(q\) is the strength of this bound, not its direction.
For \(q\) near the lower end of the window the prefactor \(1/(4q-3)\) is large and the shadow contracts sharply with \(Q\), so even a small charge pushes \(\theta_d\) out of the band and the upper bound on \(Q\) is tight.
For \(q\) near \(3/2\) the contraction is mild, \(\theta_d\) is nearly flat in \(Q\), and the bound relaxes until almost no useful limit on \(Q\) can be extracted from the size observable alone.
This monotonic, \(q\)-graded sensitivity is consistent with what is found in closely related models studied in the EHT literature \citep{Zubair:2023cep, Gogoi:2023ffh, Rincon:2023hvd, Uniyal:2022vdu}.

We construct the constraint bands as follows.
For each target and each value of \(q\) in the range Eq.~\ref{eq:sol-q-range}, we fix the spin \(a\) and the inclination \(\theta_o\), vary \(Q\) from zero up to the charge at which the configuration becomes extremal (the boundary \(a_{\rm ext}(Q,q)\) , see Fig.~\ref{fig:phase_diagram}), compute \(\theta_d\) via Eq.~\ref{eq:angdia} and Eq.~\ref{eq:areal-radius}, and compare against the measured value together with its \(1\sigma\) and \(2\sigma\) uncertainties.
At the \(1\sigma\) level the M87\(^*\) band is \(39 \leq \theta_d \leq 45\,\mu\mathrm{as}\) and the Sgr A\(^*\) band is \(41.7 \leq \theta_d \leq 55.7\,\mu\mathrm{as}\).
We define the \(1\sigma\) allowed region in the \((Q, q)\) plane as the set of parameter pairs for which the predicted \(\theta_d\) falls inside the band of a given source; the two sources are treated separately, since each black hole carries its own charge.
The \(2\sigma\) region is defined analogously using the doubled uncertainty intervals, giving a more conservative and broader allowed domain.
We carry out this analysis at the representative spins \(a\in\{0.7,\,0.9\}\,M\), to quantify how the constraint changes with rotation.
Because the shadow area shrinks with \(a\) at fixed \(Q\) (Fig.~\ref{fig:shadow_radius}), the predicted \(\theta_d\) starts closer to the lower band edge and exits it at a smaller charge, so increasing the spin shifts the upper bound on \(Q\) to smaller values at every \(q\).
The smaller extremal charge at higher spin compounds this tightening.

We map these measurements onto the \((Q, q)\) plane at fixed spin and overlay the \(1\sigma\) band of each source.
At the Maxwell point \(q = 1\) and for \(a = 0.7\,M\) the M87\(^*\) data restrict \(Q \lesssim 0.52\,M\) at \(1\sigma\) (see Tab.~\ref{tbl:eht_constraints_summary}), while the Sgr A\(^*\) size bound permits \(Q \lesssim 0.71\,M\) at the same spin.
The \(q\) ordering established above is recovered numerically and is not an artefact of measuring \(Q\) in units of \(M^{2-1/q}\): expressed as a fraction of the extremal charge, the M87\(^*\) bound rises from \(0.40\,Q_{\rm ext}\) at \(q=0.8\) to \(0.73\,Q_{\rm ext}\) at \(q=1\) and reaches \(Q_{\rm ext}\) for \(q\gtrsim1.25\), where the size alone leaves \(Q\) unconstrained beyond horizon existence.
We observe that both sources are consistent with vanishing power-Yang-Mills charge, and neither requires it.

\begin{table}
\caption{Upper bounds \(Q_{\max}\) on the Yang-Mills charge at \(1\sigma\) from
the M87\(^*\) and Sgr A\(^*\) EHT shadow diameters at fixed spin
\(a=0.7\,M\), in units of \(M^{2-1/q}\) and as fractions of the extremal
charge \(Q_{\rm ext}\) at the same spin. For each \(q\) the bound is the
charge at which the predicted diameter \(d_{\rm sh}=2R_s\) exits the
\(1\sigma\) band from below, or at which the black hole becomes
over-extremal; a ratio of \(1.00\) means that the size measurement adds
nothing beyond horizon existence.}\label{tbl:eht_constraints_summary}
\centering
\begin{tabular}{cccccc}
\hline\noalign{\smallskip}
  \(q\) & \(Q_{\rm ext}/M^{2-1/q}\) & \(Q_{\max}^{\mathrm{M87}^*}/M^{2-1/q}\) & \(Q_{\max}^{\mathrm{M87}^*}/Q_{\rm ext}\) & \(Q_{\max}^{\mathrm{Sgr\,A}^*}/M^{2-1/q}\) & \(Q_{\max}^{\mathrm{Sgr\,A}^*}/Q_{\rm ext}\) \\
\noalign{\smallskip}\hline\noalign{\smallskip}
  0.80 & 0.2787 & 0.1121 & 0.40 & 0.1961 & 0.70 \\
  0.90 & 0.5451 & 0.3109 & 0.57 & 0.4955 & 0.91 \\
  1.00 & 0.7141 & 0.5235 & 0.73 & 0.7141 & 1.00 \\
  1.10 & 0.8384 & 0.7405 & 0.88 & 0.8384 & 1.00 \\
  1.25 & 0.9791 & 0.9791 & 1.00 & 0.9791 & 1.00 \\
  1.40 & 1.0868 & 1.0868 & 1.00 & 1.0868 & 1.00 \\
\noalign{\smallskip}\hline
\end{tabular}
\end{table}

\begin{figure}
\centering
\pandocbounded{\includegraphics[keepaspectratio,alt={Angular shadow diameter \textbackslash theta\_d (in \textbackslash muas) in the (Q/M\^{}\{2-1/q\}, q) plane at a=0.7\textbackslash,M for Sgr A\^{}* (left, \textbackslash theta\_o=50\^{}\textbackslash circ) and M87\^{}* (right, \textbackslash theta\_o=17\^{}\textbackslash circ). Dashed and loosely dashed contours mark the lower edges of the 1\textbackslash sigma and 2\textbackslash sigma shadow diameter bands of each source (d\_\textbackslash mathrm\{sh\}=10.2\textbackslash pm1.5\textbackslash,M for Sgr A\^{}*, 11.0\textbackslash pm1.5\textbackslash,M for M87\^{}*), and the allowed region lies to the left of the dashed contour. The central values and upper band edges exceed the predicted \textbackslash theta\_d throughout the domain and therefore do not appear.}]{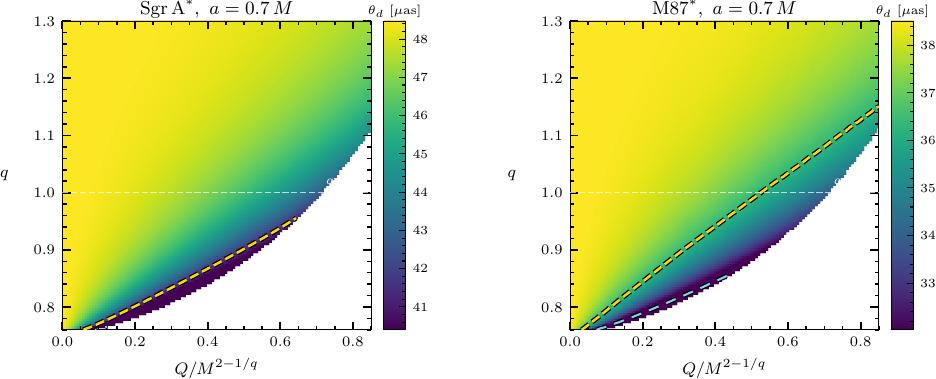}}
\caption{Angular shadow diameter \(\theta_d\) (in \(\mu\)as) in the \((Q/M^{2-1/q}, q)\) plane at \(a=0.7\,M\) for Sgr A\(^*\) (left, \(\theta_o=50^\circ\)) and M87\(^*\) (right, \(\theta_o=17^\circ\)). Dashed and loosely dashed contours mark the lower edges of the \(1\sigma\) and \(2\sigma\) shadow diameter bands of each source (\(d_\mathrm{sh}=10.2\pm1.5\,M\) for Sgr A\(^*\), \(11.0\pm1.5\,M\) for M87\(^*\)), and the allowed region lies to the left of the dashed contour. The central values and upper band edges exceed the predicted \(\theta_d\) throughout the domain and therefore do not appear.}\label{fig:eht_constraints}
\end{figure}

To exhibit the full dependence of the angular diameter on \((Q, q)\) we compute \(\theta_d\) on a two-dimensional grid for each source and display the result as a colour map. Fig.~\ref{fig:eht_constraints} presents the two resulting panels for Sgr A\(^*\) (left, \(\theta_o=50^\circ\)) and M87\(^*\) (right, \(\theta_o=17^\circ\)), each showing the \((Q/M^{2-1/q}, q)\) plane at the fixed spin \(a=0.7\,M\) used in Tab.~\ref{tbl:eht_constraints_summary}.
Overlaid on each map are the lower edges of the \(1\sigma\) (dashed) and \(2\sigma\) (loosely dashed) shadow diameter bands.
Since the charge term only shrinks the shadow, the predicted \(\theta_d\) at this spin lies below the observed central value throughout the domain, so the central and upper edge levels never intersect the displayed region and only the lower edges constrain the parameters.
Both panels make the \(q\) ordering visible at a glance, near \(q = 3/4\) the \(1\sigma\) contours close at small \(Q\), while toward the upper end of the displayed range they flatten and the size bound on \(Q\) loosens.
At \(q=1\) the dashed contours cross at \(Q \simeq 0.52\,M\) (M87\(^*\)) and \(Q \simeq 0.71\,M\) (Sgr A\(^*\)), reproducing the bounds of Tab.~\ref{tbl:eht_constraints_summary}.
Both sources deliver consistent constraints; the tighter M87\(^*\) band reflects the smaller fractional uncertainty of that measurement.

\begin{figure}
\centering
\pandocbounded{\includegraphics[keepaspectratio,alt={Allowed region of the rotating EPYM black hole in the (Q/M\^{}\{2-1/q\},q) plane at fixed spin a=0.9\textbackslash,M, with q spanning the physical window 3/4\textbackslash le q\textless3/2. The hatched grey region is the naked-singularity domain (a\_\{\textbackslash rm ext\}(Q,q)\textless0.9\textbackslash,M). The solid curve is the extremal boundary a\_\{\textbackslash rm ext\}=0.9\textbackslash,M and the dashed curve is the lower edge of the M87\^{}* 1\textbackslash sigma shadow diameter band (d\_\textbackslash mathrm\{sh\}=11.0\textbackslash pm1.5\textbackslash,M). The white strip between these two curves is a black hole excluded by the M87\^{}* measurement; the blue region satisfies both horizon existence and the M87\^{}* constraint. The dotted line marks the Maxwell point q=1.}]{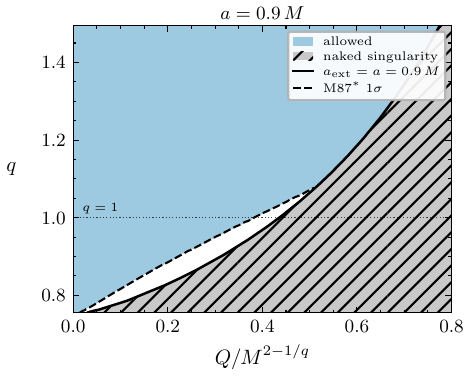}}
\caption{Allowed region of the rotating EPYM black hole in the \((Q/M^{2-1/q},q)\) plane at fixed spin \(a=0.9\,M\), with \(q\) spanning the physical window \(3/4\le q<3/2\). The hatched grey region is the naked-singularity domain (\(a_{\rm ext}(Q,q)<0.9\,M\)). The solid curve is the extremal boundary \(a_{\rm ext}=0.9\,M\) and the dashed curve is the lower edge of the M87\(^*\) \(1\sigma\) shadow diameter band (\(d_\mathrm{sh}=11.0\pm1.5\,M\)). The white strip between these two curves is a black hole excluded by the M87\(^*\) measurement; the blue region satisfies both horizon existence and the M87\(^*\) constraint. The dotted line marks the Maxwell point \(q=1\).}\label{fig:allowed_region}
\end{figure}

The likelihood analysis below places a Gaussian prior on the spin, centered at \(a=0.90\,M\), so it is natural to display the full set of restrictions at this spin in a single map. Fig.~\ref{fig:allowed_region} divides the \((Q/M^{2-1/q}, q)\) plane, with \(q\) confined to the physical window Eq.~\ref{eq:sol-q-range}, into three regions.
Where \(a_{\rm ext}(Q,q) < 0.9\,M\) no horizon exists and the singularity is naked (cf.Fig.~\ref{fig:phase_diagram}).
Inside the black-hole region we require the predicted diameter \(d_\mathrm{sh}\) to fall within the M87\(^*\) \(1\sigma\) band.
For \(q\lesssim1.07\) the M87\(^*\) \(1\sigma\) edge closes the allowed region before extremality is reached, at the Maxwell point we obtain \(Q\lesssim0.38\,M\) against the extremal charge \(Q\simeq0.44\,M\).
At larger \(q\) the extremal boundary binds first and the size measurement adds nothing beyond horizon existence.

\begin{figure}
\centering
\pandocbounded{\includegraphics[keepaspectratio,alt={Fractional Schwarzschild deviation \textbackslash delta Eq.~ as a function of the power-Yang-Mills charge Q/M\^{}\{2-1/q\} at fixed spin a=0.9\textbackslash,M, for three power exponents q\textbackslash in\textbackslash\{0.8,1.0,1.3\textbackslash\}. Left: comparison with the Sgr,A\^{}* EHT constraint (\textbackslash theta\_o=50\^{}\textbackslash circ); right: M87\^{}* (\textbackslash theta\_o=17\^{}\textbackslash circ). Green bands mark the 1\textbackslash sigma (dark) and 2\textbackslash sigma (light) intervals measured by the EHT. Increasing Q drives \textbackslash delta negative; decreasing q steepens the descent.}]{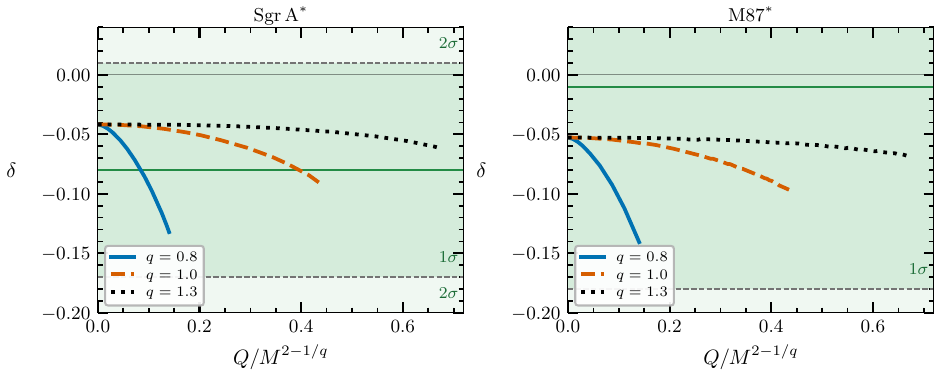}}
\caption{Fractional Schwarzschild deviation \(\delta\) Eq.~\ref{eq:eht-delta} as a function of the power-Yang-Mills charge \(Q/M^{2-1/q}\) at fixed spin \(a=0.9\,M\), for three power exponents \(q\in\{0.8,1.0,1.3\}\). Left: comparison with the Sgr,A\(^*\) EHT constraint (\(\theta_o=50^\circ\)); right: M87\(^*\) (\(\theta_o=17^\circ\)). Green bands mark the \(1\sigma\) (dark) and \(2\sigma\) (light) intervals measured by the EHT. Increasing \(Q\) drives \(\delta\) negative; decreasing \(q\) steepens the descent.}\label{fig:delta_constraints}
\end{figure}

The deviation parameter \(\delta\) Eq.~\ref{eq:eht-delta} gives a direct read on how far the shadow departs from the Schwarzschild prediction.
Because the EPYM charge term is repulsive it shrinks the photon orbit and drives \(\delta\) to increasingly negative values as \(Q\) grows. Fig.~\ref{fig:delta_constraints} displays \(\delta(Q)\) at fixed spin \(a=0.9\,M\) for three representative power exponents \(q\in\{0.8, 1.0, 1.3\}\) and superimposes the \(1\sigma\) and \(2\sigma\) EHT bands for both sources.
All three curves decrease monotonically from the Kerr baseline at \(Q=0\); the slope steepens as \(q\) decreases toward \(3/4\) because the prefactor \(2^{q-1}/(4q-3)\) amplifies the charge contribution to the lapse.
The \(1\sigma\) M87\(^*\) band, \(-0.18\lesssim\delta\lesssim0.16\), follows from the deviation \(\delta = -0.01 \pm 0.17\) obtained in the M87\(^*\) metric tests \citep{EventHorizonTelescope:2021dqv}, and contains all three curves up to extremality.
The tighter Sgr A\(^*\) band, \(-0.17\lesssim\delta\lesssim0.01\) \citep{EventHorizonTelescope:2022xqj}, also contains all three curves up to extremality, so \(\delta\) acts as a consistency check rather than an independent bound.
Both panels are mutually consistent and confirm that the sensitivity to charge, graded by \(q\), seen in the angular-diameter analysis carries over to the deviation observable.

The band construction above evaluates the constraint at a few representative spins and keeps every \(Q\) that falls within the band for at least one of them.
To strengthen the bound statistically, we carry out a likelihood analysis in which the spin is a nuisance parameter rather than a fixed input.
For each source we define

\begin{equation}\protect\phantomsection\label{eq:chi2def}{\chi^2(a, Q) = \left[\frac{\theta_d(a, Q) - \theta_{\rm obs}}{\sigma}\right]^2,}\end{equation}

where \(\theta_d\) is the model angular diameter from Eq.~\ref{eq:angdia} computed at the inclination of the corresponding source, and \(\sigma\) is the \(1\sigma\) measurement uncertainty (\(3\,\mu\)as for M87\(^*\) and \(7\,\mu\)as for Sgr A\(^*\)).
The spin is assigned a Gaussian prior \(\pi(a)\) centered on the measured values, \(a = 0.90 \pm 0.05\) for M87\(^*\) from the twisted light analysis \citep{Tamburini:2019vrf} and \(a = 0.90 \pm 0.06\) for Sgr A\(^*\) from the outflow method \citep{Daly:2023axh}, and is integrated out,

\begin{equation}\protect\phantomsection\label{eq:chi2marg}{\mathcal{L}(Q) \propto \int \exp\!\left[-\tfrac{1}{2}\chi^2(a, Q)\right]\pi(a)\, da, \qquad \Delta\chi^2(Q) = -2\ln\mathcal{L}(Q),}\end{equation}

at fixed power \(q\).
Over-extremal combinations of \(a\) and \(Q\) cast no shadow and carry zero likelihood, so the extremality boundary \(a_{\rm ext}(Q,q)\) ofSec.~\ref{sec:solution} enters the marginalization automatically.

\begin{figure}
\centering
\pandocbounded{\includegraphics[keepaspectratio,alt={Spin marginalized profile \textbackslash Delta\textbackslash chi\^{}2(Q) of the Yang-Mills charge Q from the EHT shadow diameters of M87\^{}* (blue dotted) and Sgr A\^{}* (red dashed) at the Maxwell point q=1. Dotted horizontal lines mark the 1\textbackslash sigma and 2\textbackslash sigma levels (\textbackslash Delta\textbackslash chi\^{}2=1,\textbackslash,4, one degree of freedom).}]{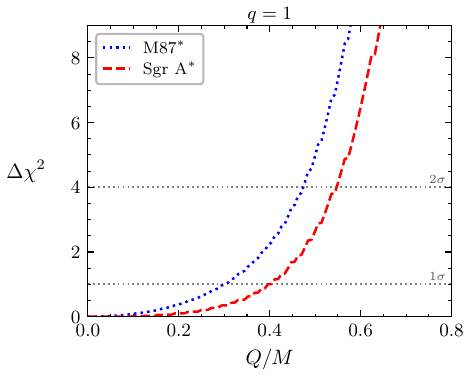}}
\caption{Spin marginalized profile \(\Delta\chi^2(Q)\) of the Yang-Mills charge \(Q\) from the EHT shadow diameters of M87\(^*\) (blue dotted) and Sgr A\(^*\) (red dashed) at the Maxwell point \(q=1\). Dotted horizontal lines mark the \(1\sigma\) and \(2\sigma\) levels (\(\Delta\chi^2=1,\,4\), one degree of freedom).}\label{fig:eht_chi2}
\end{figure}

Fig.~\ref{fig:eht_chi2} shows the resulting profiles \(\Delta\chi^2(Q)\) together with the \(1\sigma\) and \(2\sigma\) reference levels (\(\Delta\chi^2 = 1,\,4\) for one degree of freedom).
At the Maxwell point \(q=1\) the best fit lies at \(Q \approx 0\), with individual \(1\sigma\) upper bounds of \(0.29\,M\) from M87\(^*\) and \(0.39\,M\) from Sgr A\(^*\).
The M87\(^*\) likelihood bound is roughly a factor of two tighter than the band value \(Q \lesssim 0.52\,M\) of Tab.~\ref{tbl:eht_constraints_summary} because the prior concentrates the weight near \(a = 0.90\,M\) instead of fixing the spin at the representative value \(0.7\,M\).
At that higher spin the shadow contraction and the lower extremal boundary both work in the same direction.
Obviously, both sources remain consistent with vanishing charge at every \(q\).

We close with the limitations of these bounds.
The shadow diameters are evaluated at the central values of \(M\) and \(d\), so the quoted intervals do not carry the mass and distance uncertainties, for M87\(^*\) the mass error alone is at the \(\sim 10\%\) level, while for Sgr A\(^*\) the mass-to-distance ratio is tightly known from the stellar orbits \citep{Abuter:2018uum}.
Only the statistical EHT uncertainties are propagated, and systematics from the accretion model and from sparse-array image reconstruction can be comparable in size \citep{Gralla:2019xty}.
And for M87\(^*\) the measured emission ring diameter stands in for the shadow diameter, a correspondence accurate to roughly ten percent \citep{EventHorizonTelescope:2019dse, Gralla:2019xty}.

With these caveats, we adopt the M87\(^*\) \(1\sigma\) likelihood interval of Fig.~\ref{fig:eht_chi2} as the primary EHT bound on the Yang-Mills charge.

\section{Quasinormal modes}\label{sec:qnm}

We study the response of the rotating geometry Eq.~\ref{eq:metric} to a massless test scalar field.
Teukolsky demonstrated that scalar, vector, and tensor perturbations of the Kerr spacetime obey a single master equation which separates for every spin weight \citep{Teukolsky:1972my}.
The same separation is available for the Newman-Janis class at hand.
Chen and Chen showed that both the Hamilton-Jacobi and the Klein-Gordon equations separate whenever the conformal factor \(\rho^2 = r^2+a^2\cos^2\theta\) is additively separable in \(r\) and \(\theta\) and the ratio of the temporal and radial seed functions depends only on \(r\) \citep{Chen:2019jbs}, conditions which the metric Eq.~\ref{eq:metric} satisfies with its single lapse \(f(r)\).
We therefore decompose the scalar wave as \citep{Yang:2012he, Yang:2021zqy, Lambiase:2024lvo}

\begin{equation}\protect\phantomsection\label{eq:qnm-decomp}{u(t,r,\theta,\phi) = e^{-i\omega t}\,e^{i m \phi}\,u_r(r)\,u_\theta(\theta),}\end{equation}

where \(m\) denotes the azimuthal number and \(\omega\) the complex quasinormal frequency.
The quasinormal boundary conditions require waves that are purely ingoing at the outer horizon and purely outgoing at spatial infinity, selecting a discrete spectrum labelled by the multipole index \(\ell\), the azimuthal number \(m\), and the overtone \(n\) \citep{Kokkotas:1999bd, Berti:2009kk, Konoplya:2011qq}.
We write \(\omega = \omega_R + i\,\omega_I\), and \(\omega_I < 0\) means the perturbation decays.
Following the rotating Einstein-Euler-Heisenberg analysis of \citep{Lambiase:2024lvo}, we obtain the spectrum from the WKB approximation at leading order.
Higher-order WKB schemes are well developed for static backgrounds \citep{Iyer:1986np, Konoplya:2003ii, Matyjasek:2017psv, Konoplya:2019hlu}, but for rotating black holes the method becomes too involved beyond the leading corrections, so we restrict to the leading order throughout.

At the order relevant for \(\ell \gg 1\), the angular function obeys

\begin{equation}\protect\phantomsection\label{eq:qnm-angular}{\frac{1}{\sin\theta}\frac{d}{d\theta}\Big(\sin\theta\,\frac{du_\theta}{d\theta}\Big) + \Big(a^2\omega^2\cos^2\theta - \frac{m^2}{\sin^2\theta} + \mathcal{A}_{\ell m}\Big)u_\theta = 0,}\end{equation}

where \(\mathcal{A}_{\ell m}\) is the angular eigenvalue.
Since \(\omega\) is complex, \(\mathcal{A}_{\ell m}\) must be complex as well.
Its real part follows from a Bohr-Sommerfeld quantization of the angular potential between the turning points, and its imaginary part from perturbation theory of the eigenvalue problem \citep{Yang:2012he, Yang:2021zqy}.
Expanding the Bohr-Sommerfeld condition in \(a\omega/(\ell+1/2)\) gives the compact approximation

\begin{equation}\protect\phantomsection\label{eq:qnm-alm}{\mathcal{A}_{\ell m} \approx \Big(\ell+\frac{1}{2}\Big)^{2} - \frac{a^2\omega^2}{2}\left[1 - \frac{m^2}{(\ell+1/2)^2}\right],}\end{equation}

which is the form we use below.
At \(a=0\) it reduces to \((\ell+1/2)^2\), the eikonal counterpart of the spherical eigenvalue \(\ell(\ell+1)\).

The radial function satisfies a Schrödinger-type equation in the tortoise coordinate defined by \(d/dr_* = [\Delta/(r^2+a^2)]\,d/dr\),

\begin{equation}\protect\phantomsection\label{eq:qnm-radial}{\frac{d^2u_r}{dr_*^2} + V^r(r,\omega)\,u_r = 0,}\end{equation}

in which the effective potential reads

\begin{equation}\protect\phantomsection\label{eq:qnm-vr}{V^r(r,\omega) = \frac{\big[\omega(r^2+a^2) - m a\big]^2 - \Delta\,\big[\mathcal{A}_{\ell m} + a^2\omega^2 - 2 m a \omega\big]}{(r^2+a^2)^2},}\end{equation}

where \(\Delta(r)\) is the rotating radial function of Eq.~\ref{eq:sol-delta-explicit}, which carries the whole dependence on the charge \(Q\) and the exponent \(q\).
In writing Eq.~\ref{eq:qnm-vr} we kept only the terms consistent with the scalings \(\omega_R\sim\mathcal{O}(\ell)\), \(\omega_I\sim\mathcal{O}(1)\), and \(m\sim\mathcal{O}(\ell)\), so the perturbing field\textquotesingle s spin weight drops out at this order \citep{Yang:2012he, Lambiase:2024lvo}.
At \(a=0\) the tortoise coordinate reduces to \(dr_*/dr=1/f\) and the potential to \(\omega^2 - f(r)\,(\ell+1/2)^2/r^2\), the standard single-peak eikonal barrier of the static problem \citep{Kokkotas:1999bd, Berti:2009kk}.

A solution with the prescribed asymptotic behaviour exists when \(V^r\) nearly vanishes at some \(r = r_0\) and stays positive on both sides, which is where the two WKB branches are matched \citep{Iyer:1986np, Yang:2012he}.
At the first two WKB orders the matching requires

\begin{equation}\protect\phantomsection\label{eq:qnm-wkb-cond}{V^r(r_0,\omega_R) = 0, \qquad \left.\frac{\partial V^r}{\partial r}\right|_{(r_0,\,\omega_R)} = 0.}\end{equation}

These two conditions determine the pair \((r_0,\omega_R)\).
For the EPYM function \(\Delta\) a closed form in terms of radicals is not available at general \(q\), so we solve Eq.~\ref{eq:qnm-wkb-cond} numerically for each \((Q,q,a)\), seeding the root search with the Schwarzschild values \(r_0=3M\) and \(\omega_R=(\ell+1/2)/(3\sqrt{3}M)\).
Parameter points for which \(\Delta\) has no real positive root describe no horizon and are discarded.
In the static limit the conditions collapse to the photon-sphere relation \(2f(r_c)=r_c f'(r_c)\) together with \(\omega_R = (\ell+1/2)\sqrt{f(r_c)}/r_c\), so the oscillation frequency is the inverse of the static shadow radius, the familiar eikonal correspondence between quasinormal modes and the unstable photon orbit \citep{Cardoso:2008bp, Jusufi:2019ltj}.
For \(a\neq0\) the extremum \(r_0\) tracks the spherical photon orbits ofSec.~\ref{sec:shadow} and the correspondence with the shadow edge carries over in the form established for Kerr, verified there to better than \(1\%\) \citep{Yang:2021zqy}.

The imaginary part follows from the curvature of \(V^r\) at \(r_0\) through the matching of Iyer and Will \citep{Iyer:1986np},

\begin{equation}\protect\phantomsection\label{eq:qnm-omega-im}{\omega_I = -\Big(n+\frac{1}{2}\Big)\,\left.\frac{\sqrt{2\,d^2V^r/dr_*^2}}{\partial V^r/\partial\omega}\right|_{(r_0,\,\omega_R)},}\end{equation}

evaluated at the solution of Eq.~\ref{eq:qnm-wkb-cond}, and the ringdown damping rate is \(-\omega_I\).
This leading-order expression is not accurate enough for waveform comparisons with observed signals, but it captures the parameter dependence of the decay rate, which is what we need here \citep{Lambiase:2024lvo}.
We obtain \(\omega_I<0\) for every \((Q,q,a)\) explored below, so the perturbations of a massless test scalar decay within this approximation; this does not by itself establish the linear stability of the spacetime, which would require coupled gravitational and gauge-field perturbations.
Note that at \(a=0\) the expression Eq.~\ref{eq:qnm-omega-im} reduces to \(\omega_I = -(n+1/2)\,|\lambda_L|\), with \(\lambda_L\) the Lyapunov exponent associated with the unstable photon orbit \citep{Cardoso:2008bp}.
To test the leading-order WKB values at the low multipoles displayed below, we also integrate the separated radial equation directly, with the exact spheroidal eigenvalue, ingoing boundary conditions at \(r_+\), and outgoing ones at infinity.
This integration reproduces Leaver\textquotesingle s continued-fraction frequencies for Kerr \citep{Leaver:1985ax} to \(10^{-12}\), and Tab.~\ref{tbl:qnm_benchmark} shows that the WKB values differ from it by at most \(0.6\%\) at \(\ell=2\), including spins within \(1\%\) of extremality, and by about \(1.7\%\) at \(\ell=1\).

\begin{table}
\caption{Fundamental (\(n=0\)) scalar quasinormal frequencies from leading-order
WKB, \(\omega_{\rm WKB}\), and from direct integration of the radial
equation, \(\omega\). The columns \(\delta\) give
\(100\,(\omega_{\rm WKB}/\omega-1)\) for the real and imaginary parts;
for Kerr (\(Q=0\)) the last column is the distance from Leaver's
continued fraction. The rows at \(a=0.74\,M\) (\(q=0.9\)) and
\(a=0.86\,M\) (\(q=1\)) lie within \(1\%\) of the extremal spin.}\label{tbl:qnm_benchmark}
\centering
\begin{tabular}{ccccccccc}
\hline\noalign{\smallskip}
  \(q\) & \(Q/M^{2-1/q}\) & \(a/M\) & \((\ell,m)\) & \(\omega_{\rm WKB}M\) & \(\omega M\) (direct) & \(\delta\,\mathrm{Re}\,\omega\) {[}\%{]} & \(\delta\,\mathrm{Im}\,\omega\) {[}\%{]} & \(|\omega-\omega_{\rm Leaver}|M\) \\
\noalign{\smallskip}\hline\noalign{\smallskip}
  1.00 & 0.0 & 0.00 & \((2,1)\) & \(0.4811 - 0.0962\,i\) & \(0.4836 - 0.0968\,i\) & -0.52 & -0.55 & \(4\times10^{-12}\) \\
  1.00 & 0.0 & 0.50 & \((2,1)\) & \(0.5322 - 0.0935\,i\) & \(0.5354 - 0.0940\,i\) & -0.59 & -0.47 & \(3\times10^{-12}\) \\
  1.00 & 0.0 & 0.90 & \((2,1)\) & \(0.6189 - 0.0770\,i\) & \(0.6218 - 0.0772\,i\) & -0.46 & -0.26 & \(2\times10^{-12}\) \\
  1.00 & 0.0 & 0.99 & \((2,1)\) & \(0.6600 - 0.0595\,i\) & \(0.6595 - 0.0597\,i\) & +0.08 & -0.38 & \(6\times10^{-13}\) \\
  1.00 & 0.5 & 0.50 & \((2,1)\) & \(0.5647 - 0.0931\,i\) & \(0.5681 - 0.0935\,i\) & -0.60 & -0.42 & -- \\
  0.90 & 0.5 & 0.50 & \((2,1)\) & \(0.6338 - 0.0969\,i\) & \(0.6374 - 0.0973\,i\) & -0.56 & -0.36 & -- \\
  1.20 & 0.5 & 0.50 & \((2,1)\) & \(0.5388 - 0.0925\,i\) & \(0.5421 - 0.0929\,i\) & -0.60 & -0.45 & -- \\
  1.40 & 0.5 & 0.50 & \((2,1)\) & \(0.5340 - 0.0929\,i\) & \(0.5372 - 0.0933\,i\) & -0.60 & -0.45 & -- \\
  0.90 & 0.5 & 0.74 & \((2,1)\) & \(0.7350 - 0.0666\,i\) & \(0.7346 - 0.0668\,i\) & +0.05 & -0.35 & -- \\
  1.00 & 0.5 & 0.86 & \((2,1)\) & \(0.6756 - 0.0609\,i\) & \(0.6748 - 0.0612\,i\) & +0.12 & -0.57 & -- \\
  0.90 & 0.5 & 0.50 & \((1,1)\) & \(0.4127 - 0.0954\,i\) & \(0.4195 - 0.0964\,i\) & -1.62 & -1.10 & -- \\
  1.00 & 0.5 & 0.50 & \((1,1)\) & \(0.3625 - 0.0922\,i\) & \(0.3689 - 0.0934\,i\) & -1.73 & -1.22 & -- \\
\noalign{\smallskip}\hline
\end{tabular}
\end{table}

\begin{figure}
\centering
\pandocbounded{\includegraphics[keepaspectratio,alt={Leading-order WKB quasinormal frequencies of the fundamental scalar mode of the rotating EPYM black hole against the multipole index \textbackslash ell, at fixed Q=0.5\textbackslash,M\^{}\{2-1/q\}, a=0.5\textbackslash,M, m=1, n=0, for four values of the power-Yang-Mills exponent q. Left: oscillation frequency \textbackslash mathrm\{Re\}\textbackslash,\textbackslash omega. Right: damping rate -\textbackslash mathrm\{Im\}\textbackslash,\textbackslash omega.}]{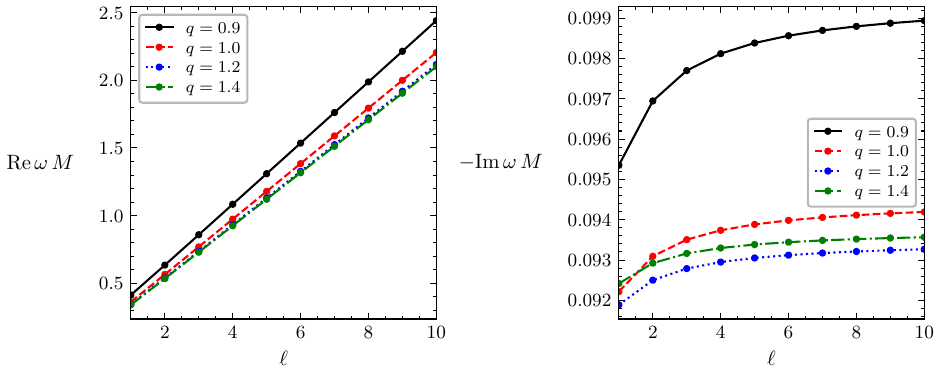}}
\caption{Leading-order WKB quasinormal frequencies of the fundamental scalar mode of the rotating EPYM black hole against the multipole index \(\ell\), at fixed \(Q=0.5\,M^{2-1/q}\), \(a=0.5\,M\), \(m=1\), \(n=0\), for four values of the power-Yang-Mills exponent \(q\). Left: oscillation frequency \(\mathrm{Re}\,\omega\). Right: damping rate \(-\mathrm{Im}\,\omega\).}\label{fig:qnm_multipole}
\end{figure}

\begin{figure}
\centering
\pandocbounded{\includegraphics[keepaspectratio,alt={Leading-order WKB quasinormal frequencies of the fundamental scalar mode (\textbackslash ell=2, m=1, n=0) against the charge Q/M\^{}\{2-1/q\}, at fixed spin a=0.5\textbackslash,M, for four values of the power-Yang-Mills exponent q. Left: oscillation frequency \textbackslash mathrm\{Re\}\textbackslash,\textbackslash omega. Right: damping rate -\textbackslash mathrm\{Im\}\textbackslash,\textbackslash omega. Each curve terminates where the configuration becomes over-extremal and no horizon exists.}]{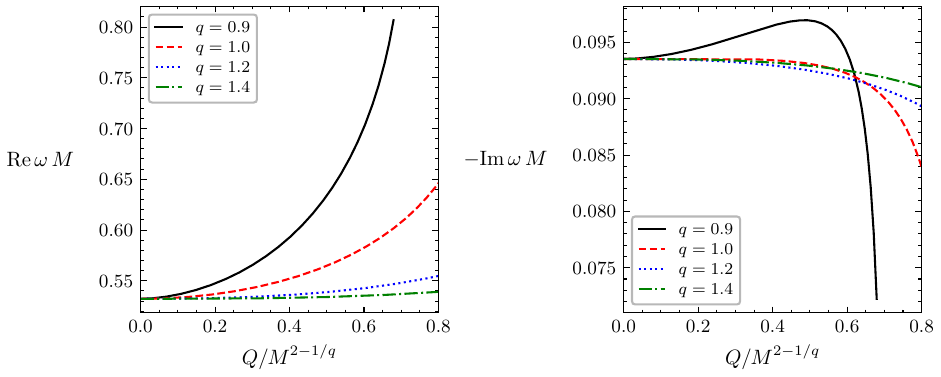}}
\caption{Leading-order WKB quasinormal frequencies of the fundamental scalar mode (\(\ell=2\), \(m=1\), \(n=0\)) against the charge \(Q/M^{2-1/q}\), at fixed spin \(a=0.5\,M\), for four values of the power-Yang-Mills exponent \(q\). Left: oscillation frequency \(\mathrm{Re}\,\omega\). Right: damping rate \(-\mathrm{Im}\,\omega\). Each curve terminates where the configuration becomes over-extremal and no horizon exists.}\label{fig:qnm_charge}
\end{figure}

\begin{figure}
\centering
\pandocbounded{\includegraphics[keepaspectratio,alt={Leading-order WKB quasinormal frequencies of the fundamental scalar mode (\textbackslash ell=2, m=1, n=0) against the spin a, at fixed charge Q=0.5\textbackslash,M\^{}\{2-1/q\}, for four values of the power-Yang-Mills exponent q. Left: oscillation frequency \textbackslash mathrm\{Re\}\textbackslash,\textbackslash omega. Right: damping rate -\textbackslash mathrm\{Im\}\textbackslash,\textbackslash omega. Each curve terminates at the extremal spin.}]{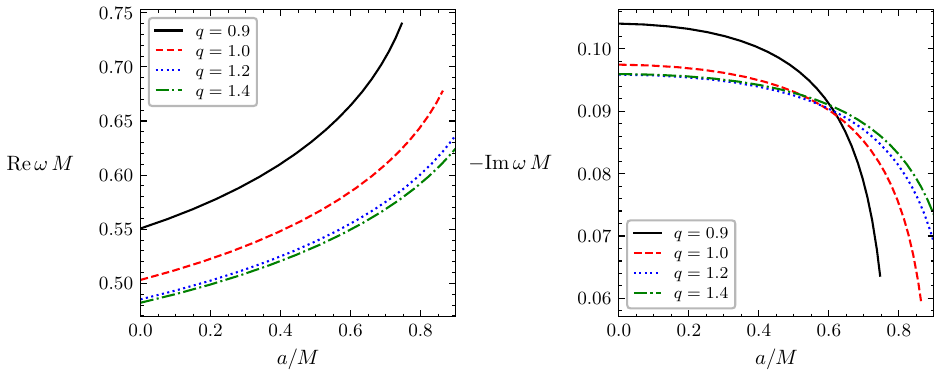}}
\caption{Leading-order WKB quasinormal frequencies of the fundamental scalar mode (\(\ell=2\), \(m=1\), \(n=0\)) against the spin \(a\), at fixed charge \(Q=0.5\,M^{2-1/q}\), for four values of the power-Yang-Mills exponent \(q\). Left: oscillation frequency \(\mathrm{Re}\,\omega\). Right: damping rate \(-\mathrm{Im}\,\omega\). Each curve terminates at the extremal spin.}\label{fig:qnm_spin}
\end{figure}

We display the fundamental mode (\(n=0\), \(m=1\)) inFig.~\ref{fig:qnm_multipole},Fig.~\ref{fig:qnm_charge}, andFig.~\ref{fig:qnm_spin}, obtained from Eq.~\ref{eq:qnm-wkb-cond} and Eq.~\ref{eq:qnm-omega-im}.
The oscillation frequency grows linearly with \(\ell\) at fixed \((Q,q,a)\), while the damping rate is nearly insensitive to \(\ell\) and saturates toward its eikonal value from below (Fig.~\ref{fig:qnm_multipole}).
The charge dependence at fixed spin is shown inFig.~\ref{fig:qnm_charge}.
We observe that \(\mathrm{Re}\,\omega\) increases with \(Q\) nonlinearly at every admitted \(q\) in the window Eq.~\ref{eq:sol-q-range}, steepest at the smallest plotted exponent \(q=0.9\) and slight at \(q=1.4\), so the exponent controls the size of the charge effect.
The damping rate decreases with \(Q\) at large charge, although near the lower edge of the window it first grows mildly before the decrease takes over.
The curves terminate where the configuration becomes over-extremal and the horizon disappears.
Since the shadow radius shrinks as \(Q\) grows (Sec.~\ref{sec:shadow}), the rise of \(\mathrm{Re}\,\omega\) with \(Q\) reflects the inverse correlation between shadow size and oscillation frequency found for other power-Yang-Mills and charged black holes \citep{Rincon:2023hvd, Gogoi:2023ffh}.
Increasing the spin at fixed charge raises \(\mathrm{Re}\,\omega\) and lowers the damping rate of the prograde mode, both nonlinearly, with the decay rate dropping fastest as extremality is approached (Fig.~\ref{fig:qnm_spin}).
The same trends were reported for the rotating Einstein-Euler-Heisenberg black hole \citep{Lambiase:2024lvo}.
Longer-lived, higher frequency ringdown is therefore the joint signature of charge and spin in this family.

We convert the dimensionless frequency \(\omega\), computed in geometrized units, to physical units through

\begin{equation}\protect\phantomsection\label{eq:hz}{f_{\rm Hz} = \mathrm{Re}\,\omega\,c^3/(2\pi G M).}\end{equation}

For M87\(^*\) with \(M = 6.5\times10^9\,M_\odot\) \citep{EventHorizonTelescope:2019dse} this gives \(f_{\rm Hz} \simeq \mathrm{Re}\,\omega\times 5.0\times10^{-6}\,\mathrm{Hz}\), placing the fundamental \(\ell=2\) mode at microhertz frequencies far below the Laser Interferometer Space Antenna (LISA) band \citep{LISA:2017pwj} but consistent with order-of-magnitude estimates for supermassive black holes.
For Sgr A\(^*\) with \(M = 4.0\times10^6\,M_\odot\) \citep{EventHorizonTelescope:2022wkp} the same conversion gives \(f_{\rm Hz} \simeq \mathrm{Re}\,\omega\times 8.1\times10^{-3}\,\mathrm{Hz}\), raising the fundamental mode to \(\sim4\times10^{-3}\,\mathrm{Hz}\), inside the LISA band.
The power-Yang-Mills charge, the exponent, and the spin together shift \(\mathrm{Re}\,\omega\) by a few percent for moderate charges, growing to tens of percent close to extremality, so the physical frequency shift is of order \(10^{-7}\,\mathrm{Hz}\) in the M87\(^*\) case and \(10^{-4}\,\mathrm{Hz}\) for Sgr A\(^*\).
Such differences lie below the resolution of current gravitational-wave detectors, but they illustrate the observability in principle of the EPYM parameter imprint in future data.

\section{Conclusions}\label{sec:conclusion}

We built a rotating EPYM black hole and used it to read three strong-field observables off a single metric.
Applying the complexification-free Newman-Janis algorithm to the static EPYM seed gave the Kerr-like solution Eq.~\ref{eq:metric}, controlled by the spin \(a\), the magnetic charge \(Q\), and the power \(q\), and reducing to Kerr-Newman at \(q=1\) and to Kerr at \(Q=0\).
Its effective source Eq.~\ref{eq:sol-eff-source} is the Kerr-Newman field at \(q=1\), where the metric solves the Einstein-Yang-Mills equations exactly, while for \(q\neq1\) it is not known to arise from a power-Yang-Mills field.
The charge term in \(\Delta(r)\) moves the horizon and the photon region inward for every \(q\) in the window, with a strength set by the power: pronounced near \(q=3/4\) and slight near \(q=3/2\).

From the photon region we computed the shadow and two observables free of coordinate dependence, the areal radius and the oblateness.
Confronting the angular size with the EHT measurements of M87\(^*\) and Sgr A\(^*\) bounds the charge for each value of \(q\) in the allowed window \(\tfrac34\le q<\tfrac32\).
Both sources are consistent with vanishing charge and place an upper limit on \(Q\) at every \(q\), since the shadow contracts monotonically with charge throughout the window.
At the Maxwell point (\(q=1\), \(a=0.7\,M\)) the M87\(^*\) angular size restricts \(Q \lesssim 0.52\,M\) at \(1\sigma\), and the spin marginalized likelihood tightens this to \(Q \lesssim 0.29\,M\).
The clearest signature of the power parameter is not a sign flip but the slope of this response: the same data bound \(Q\) tightly near \(q=3/4\), where the shadow responds sharply to charge, and only weakly near \(q=3/2\), where it is nearly insensitive.
The constraint tightens with spin, because the shadow area shrinks with \(a\) at fixed charge, placing the predicted diameter closer to the lower edge of the measured band, and because the extremal charge itself drops.

The same shadow radius sets the limiting absorption cross section at high frequencies and hence the energy emission rate, a Planck type profile peaked near \(\omega\approx3T_H\).
Increasing \(Q\) lowers \(R_s\) and \(T_H\) and so suppresses the peak and extends the lifetime, while increasing \(q\) raises both and speeds the evaporation, while for a rotating hole the estimate holds only at high frequency, because greybody factors and superradiance are neglected.

We computed the quasinormal modes of a massless scalar on the rotating background from the leading-order WKB conditions on the Teukolsky type radial potential.
The oscillation frequency rises and the damping rate falls as either the charge or the spin grows, fastest near extremality, the same inverse correlation between shadow size and ringdown frequency reported for related power-Yang-Mills black holes \citep{Gogoi:2023ffh, Rincon:2023hvd}.
Scaled to the source masses, the fundamental mode falls at microhertz frequencies for M87\(^*\) and at \(\sim4\times10^{-3}\) Hz for Sgr A\(^*\), inside the LISA band, although the EPYM shifts of a few percent lie below the resolution of current detectors.

Several extensions follow naturally.
A time-domain ringdown beyond the eikonal limit would test the correspondence we relied on.

\section{Data availability}\label{data-availability}

This manuscript has no associated data.
The analysis uses the published Event Horizon Telescope measurements of M87\(^*\) and Sgr A\(^*\) cited in the text, which are available in the referenced collaboration papers.

\bibliography{references.bib}

\end{document}